\documentclass[12pt]{article}

\usepackage{amsmath,amsfonts,amsbsy,amssymb,array,accents}
\usepackage{enumerate,array,latexsym,graphicx,mathrsfs,verbatim,psfrag,float}
\usepackage{bm} 
\usepackage{booktabs} 
\usepackage[usenames]{color}
\usepackage{empheq}

\usepackage[nosort]{cite}
\usepackage{chngpage} 

\usepackage{fancyhdr}

\usepackage[colorlinks=true,      linkcolor=darkblue,      urlcolor=darkblue,   linktoc=all,   
            filecolor=darkblue,      citecolor=darkblue,       pdfstartview=FitH,     
						pdfpagemode=UseNone,      bookmarksopen=true]{hyperref}  
\usepackage[all]{hypcap}     

\newcommand{\captionfonts}{\small}
\makeatletter  
\long\def\@makecaption#1#2{%
  \vskip\abovecaptionskip
  \sbox\@tempboxa{{\captionfonts #1: #2}}%
 \ifdim \wd\@tempboxa >\hsize
    {\captionfonts #1: #2\par}
  \else
    \hbox to\hsize{\hfil\box\@tempboxa\hfil}%
  \fi
  \vskip\belowcaptionskip}
\makeatother   

\DeclareMathSymbol{\medhatsym}{\mathord}{largesymbols}{"62} 

\DeclareMathSymbol{\medtildesym}{\mathord}{largesymbols}{"65}

\newcommand{\comm}[1]{} 

\def\({\left(}
\def\){\right)}
\def\[{\left[}
\def\]{\right]}

\def\be{{\begin{equation}}}
\def\ee{{\end{equation}}}

\def\One{{\hbox{ 1\kern-.8mm l}}}

\def\barray{\begin{array}}
\def\earray{\end{array}}
\def\be{\begin{equation}}
\def\ee{\end{equation}}
\def\bea{\begin{eqnarray}}
\def\eea{\end{eqnarray}}
\def\bal{\begin{align}}
\def\eal{\end{align}}
\def\nn{\nonumber}

 \numberwithin{equation}{section} 

\makeatletter
\g@addto@macro\bfseries{\boldmath}
\makeatother

\definecolor{cardinal}{rgb}{0.6,0,0}
\definecolor{darkgreen}{rgb}{0,0.4,0}
\definecolor{purple}{rgb}{0.5, 0, 0.5}
\definecolor{golden}{rgb}{0.92, 0.7, 0}
\definecolor{midnight}{rgb}{0, 0, 0.5}
\definecolor{darkblue}{rgb}{0, 0, 0.8}

\def\cA{{\cal A}}
\def\cB{{\cal B}}

\def\cE{{\cal E}}
\def\cF{{\cal F}}

\def\cH{{\cal H}}
\def\cJ{{\cal J}}
\def\cL{{\cal L}}

\def\cN{{\cal N}}
\def\cO{{\cal O}}

\def\cS{{\cal S}}

\def\cV{{\cal V}}
\def\cW{{\cal W}}

\newcommand{\sM}{{\scriptscriptstyle M}}
\newcommand{\sN}{{\scriptscriptstyle N}}
\newcommand{\sP}{{\scriptscriptstyle P}}
\newcommand{\sQ}{{\scriptscriptstyle Q}}
\newcommand{\sR}{{\scriptscriptstyle R}}
\newcommand{\fg}{\mathfrak{g}}
\newcommand{\Vol}{\mathrm{Vol}}

\def\cO{{\cal O}}

\newcommand{\im}{\mathrm{i}}
\newcommand{\?}{\;\!}

\def\Kah{\mathrm{J}}
\def\Kahf{K\"ahler }
\def\Ric{\mathrm{R}}

\newcommand{\ex}{{\mathrm{e}}}
\renewcommand\Im{\hbox{{\rm Im}}\,}
\renewcommand\Re{\hbox{{\rm Re}}\,}
\newcommand{\Iprod}[2]{\langle {#1}, {#2} \rangle}
\newcommand{\nv}{n_\mathrm{v}}
\newcommand{\cK}{\mathcal{K}}
\def\u{\mathsf{u}}

\newcommand{\ominf}{\omega_{\scriptscriptstyle \infty}}
\newcommand{\Jpar}{\mathrm{j}}
\def\scr2{{\scriptscriptstyle (2)}}
\newcommand{\qv}{\mathcal{R}}
\newcommand{\sv}{\mathcal{I}}
\newcommand{\vvv}{\mathrm{v}}
\begin{document}

\pagenumbering{Alph}

\begin{titlepage}

\begin{flushright}
{\fontsize{14pt}{0pt} CPHT -RR102.112018 }
\end{flushright}

\begin{center}

\vskip .3in \noindent

{\Large \bf{Rotating attractors and BPS black holes in AdS$_4$}}

\bigskip

	Kiril Hristov$^{1}$, Stefanos Katmadas$^{2}$ and Chiara Toldo$^{3,4}$\\

       \bigskip
       $^{1}$Institute for Nuclear Research and Nuclear Energy, \\ Bulgarian Academy of Sciences, 1784 Sofia, Bulgaria \\
 \medskip
       $^{2}$ Instituut  voor  Theoretische  Fysica,  KU  Leuven,  \\ Celestijnenlaan  200D,  B-3001  Leuven,  Belgium \\
       \medskip
	$^{3}$ Kavli Institute for Theoretical Physics, Kohn Hall, \\
	University of California Santa Barbara, CA, 93106\\
	\medskip
	$^{4}$ Centre de Physique Th\'eorique  (CPHT), Ecole Polytechnique, \\
	91128 Palaiseau Cedex, France

       \vskip .5in
       {\bf Abstract }
       \vskip .1in

       \end{center}

We analyze stationary BPS black hole solutions to $4d$ $\mathcal{N}=2$ abelian gauged supergravity. Using an appropriate near horizon ansatz, we construct rotating attractors with magnetic flux realizing a topological twist along the horizon surface, for any theory with a symmetric scalar manifold. An analytic flow to asymptotically locally AdS$_4$ is presented for a subclass of these near-horizon geometries, and an explicit new example of supersymmetric AdS$_4$ rotating black hole is discussed in detail. We further note that, upon tuning the gauging to special values, one can obtain solutions with different asymptotics, and in particular reductions of doubly spinning asymptotically AdS$_5$ black holes. Finally we present a proposal for the form of the BPS entropy function with rotation, which we expect to be holographically related to the refined twisted index of the dual theory.

\vfill
\eject

\thispagestyle{empty}

\end{titlepage}
\pagenumbering{arabic}

{\hypersetup{linkcolor=black}
\tableofcontents
}
\section{Introduction and summary of results}
\label{sec:intro}

Supersymmetry is an efficient tool for understanding the microscopic structure of black holes, as exemplified by the wealth of developments and precision tests achieved in the last two decades, building on the original constructions of \cite{Strominger:1996sh, Maldacena:1997de} for counting the entropy for asymptotically flat BPS black holes. Supersymmetric black holes in asymptotically anti-de Sitter (AdS) spacetimes with a known field theory dual provide a solid testing ground for extending these results beyond the near-horizon region and into more general examples, e.g.\ by allowing more general horizon topologies and including rotation. 

The construction of the relevant black hole solutions in four dimensions has a long history (see e.g.\ \cite{Romans:1991nq, Gauntlett:2001qs}), leading to the first examples of static supersymmetric black holes in AdS$_4$ with a spherical horizon of finite area in gauged supergravity coupled to vector multiplets \cite{Cacciatori:2009iz,Dall'Agata:2010gj,Hristov:2010ri}. These examples were later extended to allow for a full set of electromagnetic charges
\cite{Halmagyi:2013qoa, Katmadas:2014faa, Halmagyi:2014qza}. Most recently, the microstate counting for these static solutions in the gauged STU model (admitting an uplift to M-theory) was performed successfully in the large $N$ limit of the dual ABJM theory via supersymmetric localization \cite{Benini:2015eyy,Benini:2016hjo,Benini:2016rke,Cabo-Bizet:2017jsl}. The match was also extended to various other examples and in different directions, see e.g.\ \cite{Hosseini:2016tor,Azzurli:2017kxo,Liu:2017vll,Hosseini:2017fjo,Benini:2017oxt,Liu:2018bac,Hristov:2018lod} and references therein. The presence of a magnetic flux (in the electrically gauged symplectic frame) is crucial for preserving supersymmetry by the so called supersymmetric twist: the contribution of the spin connection cancels against the gauged R-symmetry in the BPS equations. The associated ground states of the dual field theory with magnetic flux are counted by the topologically twisted index \cite{Benini:2015noa}.

The twist characterizing the static BPS solutions in AdS$_4$ leads to an additional special feature: the fermionic symmetries commute with spatial rotations \cite{Hristov:2011ye}. It is therefore immediate to see that adding nonzero angular momentum does not break any further symmetries, unlike the case of supersymmetric asymptotically flat black holes. Constructing explicitly such black holes is however a nontrivial task and it is the purpose of the present work to find such solutions in $\mathcal{N}=2$ gauged supergravity models with embedding in string or M-theory. 

The supersymmetric rotating black hole solutions that have appeared in the literature so far include electrically charged 1/4-BPS solutions \cite{Kostelecky:1995ei} without a regular static limit, as well as 1/4-BPS black holes with magnetic charges \cite{Caldarelli:1998hg, Klemm:2011xw}, which must necessarily have a non-compact horizon. The known non-extremal rotating black hole configurations supported by scalars and gauge fields \cite{Chong:2004na,Chow:2010sf,Chow:2010fw,Chow:2013gba,Gnecchi:2013mja}, also display similar features and are not continuously connected to static magnetic black holes with spherical horizons, for which the counting is known.

In this paper, we construct 1/4-BPS black hole solutions that have either compact or non-compact horizons and realize the same topological twist via magnetic flux\footnote{Here we again refer to a frame where the gauging is purely electric. More generally, this requirement implies that the vector of FI gaugings must be mutually nonlocal with the vector of charges, i.e. their inner product must not vanish.} as the static black holes of \cite{Cacciatori:2009iz,Dall'Agata:2010gj,Hristov:2010ri}, to which they reduce in a static limit. We find that the rotation manifests itself as a slicing of spacetime in terms of a squashed sphere (or a squashed hyperbolic space in the non-compact case), rather than a constant curvature one as in the static case. We therefore expect the microstate counting for these rotating AdS black holes to be based on the refined topologically twisted index on a similarly squashed background as described in \cite{Benini:2015noa}. Such a microstate counting can provide precious insight on the microscopic structure of more realistic spinning black holes, belonging to the class of extremal Kerr with or without cosmological constant, since the relevant degrees of freedom are expected to be universal and reside in the near horizon geometry.

In this work we make an extensive use of the real formulation of supergravity \cite{Klemm:2012yg, Klemm:2012vm} in terms of the rank-four symplectic covariant tensor $I_4$ \cite{Ferrara:1997uz,Ferrara:2006yb,Bossard:2013oga}. Although this way of repackaging $4d$ ${\cal N}=2$ supergravity is less popular in the literature, we find it very convenient for solving the BPS equations. We therefore give an extensive introduction to the real formulation and the properties of the quartic invariant. The manifest symplectic covariance of the real formulation makes it possible to solve for a very general near-horizon ansatz, which allows for a full set of electromagnetic charges $(q_I, p^I)$, subject to the twisting condition\footnote{The vanishing of the NUT charge, imposed to ensure causality, provides an additional constraint among the set of conserved charges.}, and angular momentum ${\cal J}$ bounded from above, for arbitrary symmetric models and gaugings. The real, or $I_4$, formalism then also allows us to write down full analytic flows to the asymptotic AdS$_4$ region in the models we consider.

We finish this work with a proposal for the BPS entropy function for our rotating solutions, such that we can formulate the black hole entropy and attractor equations through an extremization principle. The existence of a BPS entropy function has proven crucial in the successful holographic matches so far. Adding rotation, we find a more intricate form of the entropy function that simplifies drastically only for a special class of models that does not exhibit AdS$_4$ asymptotics. Our results therefore provide a nontrivial check for the large $N$ evaluation of the refined twisted index of the holographically dual theories.

As an appetizer, we can already present the general formula for the Bekenstein-Hawking entropy of the spherical rotating black holes in AdS$_4$,
\begin{equation}
 S_{BH} 
= \frac{\pi l_{AdS}^{2} } { \sqrt2 G_N}  \sqrt{ \frac14\?I_4(\Gamma, \Gamma, G, G) + \sqrt{\frac1{16}\?I_4(\Gamma, \Gamma, G, G)^2 - 4  \frac{(I_4(\Gamma) + \cJ^2)}{l_{AdS}^{4}} }}\,,
\end{equation}
given in terms of the AdS length scale $l_{AdS}$, the conserved angular momentum $\cJ$, and the symplectic vectors of electromagnetic charges $\Gamma$ and gauging parameters $G$\footnote{Notice that these black holes, just like those considered for the microstate counting in the static case \cite{Cacciatori:2009iz}, do not have a well defined flat space limit.}. The rank-four symplectic tensor $I_4$ then encodes the properties of the scalar manifold, and we discuss it more explicitly in the main text emphasizing its properties for the STU model (which has a higher-dimensional origin for several choices of gauging vectors $G$).

Some other interesting questions are left for future investigations. For instance, in this paper we have constructed black hole solutions which do not admit an ergosphere associated to the Killing vector $\partial_t$: it would be interesting to construct the additional Kerr-like orbit, and ultimately connect these solutions as limits of a rotating black hole with temperature (non-extremal generalization), along the lines of \cite{Chow:2013gba,Gnecchi:2013mja}. It would be moreover interesting to compute the renormalized on-shell action of these solutions, and elucidate the holographic mapping of the fugacities. Another interesting direction would be to use the general form of the four-dimensional solution presented here to generate rotating five-dimensional black string solutions \cite{Benini:2013cda}, via the uplift procedure along the lines of \cite{Hristov:2014eza,Hosseini:2017mds,Azzola:2018sld}. We hope to come back to these points in the near future.

The rest of this paper is organized as follows. Section \ref{sec:BPSequations} provides some necessary background, in particular the real formulation of the supergravity degrees of freedom and the relevant BPS equations for models with symmetric scalar manifold. In section \ref{sec:warm-up} we summarize some results on the static solutions and their possible asymptotics that are useful in later sections and allow to build intuition on the construction of solutions in the real formulation. We then proceed to analyze the rotating near horizon geometry in section \ref{sec:nhg}, adopting an ansatz that incorporates the topological twist for a squashed internal space. This results in a family of rotating BPS attractors satisfying the same number of constraints as the corresponding static one, to which it reduces when the rotation parameter is turned off. We provide several examples for such near-horizon geometries in section \ref{subsec:nhg-examples}, including (the Kaluza-Klein reduction of) the near-horizon geometry of the five dimensional solutions in \cite{Gutowski:2004yv, Kunduri:2006ek}, as well as new examples in the STU model with gaugings that admit an AdS$_4$ vacuum.
In section \ref{sec:full-sol} we study the complete BPS black hole flow interpolating between the near-horizon geometry and asymptotically locally AdS$_4$ space, once again providing an analytic example solution. We further comment on the asymptotic properties of our solutions and identify the conserved quantities. Finally, section \ref{sec:entropy} focuses on the entropy function for our family of solutions, providing an appropriate extremization principle similar to the one used in the static case to match to the twisted index in the dual field theory. The two appendices provide more details on our conventions on $\cN=2$ supergravity coupled to vector multiplets with a symmetric scalar manifold and on the reduction of the solutions of \cite{Gutowski:2004ez,Gutowski:2004yv, Kunduri:2006ek}, respectively.

\section{Supergravity formalism and BPS equations}
\label{sec:BPSequations}

\subsection{FI gauged supergravity}
\label{subsec:sugra}

Our starting point is the bosonic action for abelian gauged supergravity \cite{deWit:1984pk, deWit:1984px}, 
\begin{equation}
\label{Ssugra4D}
S_\text{4D}=\frac{1}{16\pi}\int_{M_4} \Bigl(R\star_4 1 - 2\,g_{i\bar\jmath}\,d t^i\wedge\star_4 d \bar{t}^{\bar\jmath} - \tfrac12 F^I\wedge G_I + 2\,V_g\,\star_4 1\Bigr),
\end{equation}
which describes neutral complex scalars $t^i$ (belonging to the $\nv$ vector multiplets) and abelian gauge fields $F_{\mu\nu}{}^I$, $I=\{0,\,i \}=0,\dots \nv$ (from both the gravity multiplet and the vector multiplets), all coupled to gravity. We refer to appendix \ref{sec:conventions} for more details on our conventions for $\cN\!=\!2$ supergravity. The dual gauge fields, $G_{\mu\nu}{}_I$, are given in terms of the field strengths and the scalar dependent period matrix $\cN_{IJ}$, by
 \begin{equation}\label{G-def}
  G^-_{\mu\nu}{}_I = \cN_{IJ} F^-_{\mu\nu}{}^J\,,
 \end{equation}
where the explicit expression for the period matrix is not necessary in the following. The metric on the special K\"ahler target space $g_{i\bar\jmath}$ and the period matrix $\cN_{IJ}$ are completely specified by a holomorphic homogeneous function of degree two $F(X)$, of the projective coordinates $X^I$ on the scalar manifold, as in \cite{deWit:1984pk, deWit:1984px}. One may recover the physical scalars as $t^i=X^i/X^0$.
One can repackage the scalars in terms of a symplectic covariant section, as
\begin{equation}\label{eq:sym-sec-0}
\cV=\ex^{{\cal K}/2} \begin{pmatrix} X^I\\ F_I\end{pmatrix}\,, \qquad F_I= \frac{\partial F}{\partial X^I}\,,
\end{equation}
where $\cK$ is the K\"ahler potential and $F_I$ stands for the derivatives of $F(X)$. Note that $\cV$ is uniquely specified by the physical scalars up to a local U(1) transformation.

Finally, the scalar potential $V_g$ takes the form
\begin{equation}\label{gau-pot}
V_g= g^{i\bar\jmath}\?Z_i(G)\,\bar Z_{\bar\jmath}(G) -3\,\left|Z(G)\right|^2 \,, 
\end{equation}
where we used the definitions of the central charge
\begin{align}\label{ch-def-0}
Z(G) = \Iprod{G}{\cV} \equiv \ex^{{\cal K}/2} \left( g_I\? X^I - g^I\? F_I \right)  \,,
\end{align}
and its K\"ahler covariant derivative, $Z_i(G)$. The constant symplectic vector $G=\{g^I, g_I\}$ defines the so-called Fayet-Iliopoulos (FI) terms, which specialize the precise combination of $U(1)$ gauge fields gauging the R-symmetry. Note that in \eqref{ch-def-0} we defined in passing the inner product between two symplectic vectors, which will be often used in the following.

In the abelian class of gaugings we consider in this paper, the FI terms specify the coupling of the gravitini\footnote{We have omitted the fermionic completion of the Lagrangian, which can be found in \cite{deWit:1984pk, deWit:1984px}.}, as their kinetic term contains the minimal coupling
\begin{gather}\label{eq:gravitino}
\epsilon^{\mu\nu\rho\sigma} \bar\psi_{\mu}{}_i\gamma_\nu\,D_\rho\psi_{\sigma}{}^i \equiv
\epsilon^{\mu\nu\rho\sigma}
\bar\psi_{\mu}{}_i\gamma_\nu\left( \partial_\rho + \tfrac{i}2\,\Iprod{G}{A_\rho} \right)\psi_{\sigma}{}^i\,,
\\
\Iprod{G}{A_\mu} \equiv g_I A_\mu{}^I - g^I A_{\mu}{}_I\,.\nonumber
\end{gather}
This coupling is in general non-local, due to the presence of the dual gauge fields $A_{\mu}{}_I$. However, as for any vector, $G$ can always be rotated to a frame such that it is purely electric, i.e.~$g^I=0$, leading to a local coupling of the gauge fields. More generally, one can consider couplings of magnetic vectors as well, using the embedding tensor formalism \cite{deWit:2005ub, deWit:2011gk}, which requires the introduction of extra auxiliary fields. For the theories discussed in this paper however, the bosonic action is only affected through the nontrivial potential \eqref{gau-pot}, which can be straightforwardly written in an electric/magnetic covariant way, as above, reproducing the result obtained using the embedding tensor formalism. Based on this observation, we use covariant versions of all quantities, since all results for the bosonic backgrounds must necessarily be covariant under electric/magnetic duality. We therefore need not choose a frame for the FI terms explicitly from the outset, only specifying a frame when discussing examples.

The symplectic section $\cV$, along with the vector of FI terms, $G$, and the natural symplectic vectors of field strengths and charges in \eqref{eq:dual-gauge}, may be used to describe all the supergravity degrees of freedom in terms of symplectic covariant objects. The action \eqref{Ssugra4D} can be recast as
\begin{equation}
\label{Ssugra4D-sym}
S_\text{4D}=\frac{1}{16\pi}\int_{M_4} \Bigl(R\star_4 1  - \im\?\Iprod{D^\mu \bar\cV}{D_\mu \cV} - \tfrac12 F^I\wedge G_I + 2\,V_g\,\star_4 1\Bigr) \,,
\end{equation}
where $D_\mu$ stands for the K\"ahler connection. Note that the kinetic terms for the vector fields cannot be written in a manifestly duality covariant form, while keeping Lorentz invariance.

As in this paper we concentrate on stationary black hole solutions, we can in fact describe the degrees of freedom of the vector fields in a duality covariant form. The requirement of a timelike isometry is enough to cast the spacetime in the form 
\begin{equation}\label{eq:metr-bps}
ds^2_4 = -\ex^{2\?U} \?(dt+\omega )^2 + \ex^{-2\?U}\?ds^2_3 \,,
\end{equation}
where $ds^2_3$ is the metric of a three-dimensional base space, on which the function $U$, the one-form $\omega$ and all other fields discussed above are defined. Similarly, the gauge field strengths in \eqref{eq:dual-gauge} can be decomposed as 
\begin{gather}\label{eq:F-time-dec}
 \mathsf{F} = d\big( \zeta\,(dt+\omega) \big) + \cF = d\big( \zeta\,(dt+\omega) \big) + d \cA \,, 
\end{gather}
where $\zeta$, $\cF$, $\cA$ denote the symplectic vectors of timelike components and spatial field strengths and potentials, for both electric and magnetic gauge fields. The complex self-duality condition \eqref{cmplx-sdual} can be used to relate the time- and spacelike components as
\begin{equation}\label{eq:zeta-F}
 d \zeta = \ex^{2\?U} \?\mathrm{J}\?\star \left( \cF + \zeta\? d\omega \right)\,,
\end{equation}
where $\star$ will denote the three-dimensional Hodge operator for the remainder of this paper.
Finally, the degrees of freedom of the scalars are again described by the section, in a combination encoding both the physical scalars and the scale factor $\ex^U$, as
\begin{equation}\label{eq:q-def}
 \qv = 2\?\ex^{U}\? \Re \left( \ex^{-\im\?\alpha} \cV \right)\,,
\end{equation}
where $\alpha$ is an arbitrary function, parametrizing the unphysical overall phase of the symplectic section.
Using the timelike isometry one may reduce the original four-dimensional action down to three dimensions, which leads to the so-called real formulation of special geometry. Sparing the reduction details, here we directly give the resulting Lagrangian as derived in \cite{Klemm:2012yg, Klemm:2012vm} 
\begin{align} \label{eq:real-formulation-Lag} 
e^{-1} \cL = &\,\, \frac{1}{2}\? R_3 - \tilde{H}^{\sM\sN} \left(\?\partial \qv_\sM \? \partial \qv_\sN - \partial \zeta_\sM \? \partial \zeta_\sN \?\right) + \frac{1}{H}\? V_{3d} 
\nonumber \\ 
&\, - \frac{1}{16\?H^2} \? \Iprod{\qv}{ \partial \qv}^2 + \frac{1}{8\?H^2} \? \Iprod{\qv}{ \partial \zeta}^2  
 - \frac{1}{4 H^2} \left( \partial \tilde{\phi} + \tfrac12\? \Iprod{\zeta}{ \partial \zeta}  \right)^2 \,,
\end{align}
where the field strengths $\cF$ were dualised into the $\zeta$'s using \eqref{eq:zeta-F} and the scalar $\tilde{\phi}$ is defined through
\begin{equation}
 d \tilde{\phi} + \tfrac12\? \Iprod{\zeta}{ d \zeta}  = 4\?H^2 \star d \omega \,.
\end{equation}
The function $H(\qv)$ is the so-called Hesse potential, which plays a role analogous to the prepotential in four dimensions, and in fact the two objects are related through the reduction. When evaluated with $\qv$ as in \eqref{eq:q-def}, the Hesse potential evaluates to  
\begin{equation}
H = -\tfrac12\? \ex^{2\?U} \,, 
\end{equation}
while the remaining objects in \eqref{eq:real-formulation-Lag} are given as
\begin{equation}
\tilde{H} = -\frac12 \? \log( -2 H )   \,,
\qquad
\tilde{H}^{\sM} = \frac{\partial \tilde{H}}{\partial \qv_\sM}\,,
\qquad
\tilde{H}^{\sM\sN} = \frac{\partial^2 \tilde{H}}{\partial \qv_\sM \partial \qv\sN} \,.
\end{equation}
The scalar potential $V$ in \eqref{gau-pot} leads to the following expression for the potential in three dimensions
\begin{equation}\label{eq:3d-pot}
 \frac{1}{H}\?V_{3d} = G_\sM\? G_\sN\?\left( - \tilde{H}^{\sM\sN}  + 4\? \tilde{H}^{\sM} \tilde{H}^{\sN} \right)  + \frac{1}{8\?H^2}\? \Iprod{G}{\qv}^2 \ .
\end{equation}

Before closing, we point out the existence of the dual coordinates, defined as 
\begin{equation}\label{eq:dual-coo}
 \sv_\sM = -2\? \Omega_{\sM\sN}\tilde{H}^\sN = 2\? \ex^{-U}\? \Im \left( \ex^{-\im\?\alpha} \cV \right)_\sM \,,
\end{equation}
where $\Omega_{\sM\sN}$ is the (inverse) anti-symmetric symplectic structure matrix that serves for raising and lowering indices and in the second equality we re-expressed this object in terms of the section. This equation expresses the fact that only half of the components in the symplectic section \eqref{eq:sym-sec-0} are independent, so that its imaginary part is given in terms of the real part. Note that the variables $\sv$ are useful in obtaining explicit solutions, as will be seen in the following, so that an explicit algebraic way of computing the Hesse potential $H$ is very useful. This is true for the symmetric models considered in the rest of this paper.

\subsection{Symmetric models and the real formulation}
We henceforth fully restrict our considerations to models whose special K\"ahler target manifold parametrized by the scalars is a symmetric space, classified in \cite{deWit:1992wf}. For the cubic models, i.e. whose prepotential is of the form
\begin{equation}\label{prep-def-0}
F=\frac{1}{6}\?c_{ijk}\?\frac{X^i X^j X^k}{X^0} \,,
\end{equation}
this requirement translates into the following identity for the constant tensor $c_{ijk}$
\begin{equation}\label{eq:symm-cub}
 \frac43\?\delta_{i(l}\? c_{mpq)} = c_{ijk}\?c_{j^\prime (l m} \?c_{pq)k^\prime}\?\delta^{j j^\prime}\?\delta^{k k^\prime}\,.
\end{equation}
In terms of duality covariant objects, the property \eqref{eq:symm-cub} is expressed by the existence of a rank-4 symplectic tensor satisfying a number of identities, discussed in Appendix \ref{app:I4}. When contracted e.g.\ with a charge vector $\Gamma = \{p^I, q_I\}$, this defines the quartic form
\begin{eqnarray}\label{I4-ch-0}
I_4(\Gamma)= - (p^0 q_0 + p^i q_i)^2 + \frac{2}{3} \,q_0\,c_{ijk} p^i p^j p^k - \frac{2}{3} \,p^0\,c^{ijk} q_i q_j q_k 
             + c_{ijk}p^jp^k\,c^{ilm}q_lq_m\,, 
\end{eqnarray}
which is invariant under symplectic transformations. This is particularly useful in view of the fact that symplectic transformations do not necessarily leave the form of the prepotential \eqref{prep-def-0} invariant, while a prepotential might not exist at all in certain frame. The duality covariant formulation, based on the quartic invariant \eqref{I4-ch-0}, allows to treat all frames on the same footing.

The quartic invariant provides an explicit solution for the Hesse potential $H$ of symmetric models, which encodes the three-dimensional real formulation of the theory \eqref{eq:real-formulation-Lag}, as
\begin{equation}
 H = -\frac12\?\ex^{2\?U} = -\frac12\? \sqrt{I_4(\qv)} \,.
\end{equation}
This identification allows for a completely algebraic rewriting of the various objects in the previous section in terms of contractions of $I_4$ with the various symplectic vectors. In particular, \eqref{eq:dual-coo} becomes
\begin{equation}
 \sv =  \frac1{2\?I_4(\qv)}\? I^\prime_4(\qv) =  \frac12\? \ex^{-4\?U}\? I^\prime_4(\qv)\,,
\end{equation}
while its inverse is similarly given by
\begin{equation}\label{eq:r-i}
 \qv = - \frac1{2\?I_4(\sv)}\? I^\prime_4(\sv) = - \frac12\? \ex^{4\?U}\? I^\prime_4(\sv)\,.
\end{equation}
We refer again to Appendix \ref{app:I4} for the definition of $I^\prime_4$ and higher derivatives of $I_4$ as symplectic tensors, as well as various identities among them.
In what follows we will make extensive use of the quartic invariant $I_4$ contracted with the various symplectic vectors describing the scalars, vectors, charges, FI terms and so on in the theory. In fact one can completely write the Lagrangian using $I_4$ contractions. For future reference, we provide the expressions for the derivatives of $\tilde H$, from which the Lagrangian \eqref{eq:real-formulation-Lag} is built
\begin{align}
 \tilde{H}^\sM &  = -\frac{1}{4}\?\frac{ \partial^\sM I_4(\qv)}{I_4(\qv)}\,,
\\
\label{eq:Hab-tensor}
\tilde H^{\sM\sN} & = \frac{1}{4\?I_4(\qv)}\?\left( - \partial^\sM\partial^\sN I_4(\qv) + \frac{1}{I_4(\qv)}\? \partial^\sM I_4(\qv) \? \partial^\sN I_4(\qv)\right)\,.
\end{align}

A benchmark model in this class is provided by the STU model (sometimes also called electric STU model to distinguish it from its symplectically rotated version we discuss below), with prepotential
\begin{equation}\label{eq:STU-def}
F^{\scriptscriptstyle STU}= \frac{X^1 X^2 X^3}{X^0} \,,
\end{equation}
so that it describes three vector multiplets coupled to supergravity. In this case the only non-vanishing coefficients are $c_{123} = 1$ and permutations lead to $c^{123} = 1$ and permutations. The quartic invariant of a charge vector is therefore
\begin{align}\label{I4-ch-stu}
I_4(\Gamma)^{\scriptscriptstyle STU}=&\? - (p^0 q_0 + p^i q_i)^2 + 4\,q_0 p^1 p^2 p^3 - 4\,p^0 q_1 q_2 q_3 
\nonumber\\
                              &\?   + 4 (p^1 p^2 q_1 q_2 + p^1 p^3 q_1 q_3 + p^2 p^3 q_2 q_3)\,. 
\end{align}
From this expression, it is clear that any purely electric or magnetic $\Gamma$ leads to $I_4(\Gamma)^{\scriptscriptstyle STU}=0$. This in turn implies that an AdS$_4$ vacuum in this model requires a mixed electric/magnetic vector of FI terms, $G$, since the cosmological constant is controlled by $I_4(G)$ in any symmetric model (which we show carefully in section \ref{sec:warm-up}). 

If we apply the symplectic transformation defined by
\begin{equation}\label{eq:STU-sym-rot}
 \begin{pmatrix} p^0 \\ p^i \\ q_i \\ q_0  \end{pmatrix} \rightarrow \begin{pmatrix} -p^0 \\ - q_i \\ p^i \\ -q_0  \end{pmatrix}\,,
\end{equation}
on all symplectic vectors, one transforms the prepotential \eqref{eq:STU-def} into the magnetic STU model
\begin{equation}\label{eq:STU-root}
F^{\scriptscriptstyle mSTU}= 2\?i \sqrt{X^0 X^1 X^2 X^3}\,,
\end{equation}
while \eqref{I4-ch-stu} becomes
\begin{align}\label{I4-ch-root}
I_4(\Gamma)^{\scriptscriptstyle mSTU}=&\? - (p^0 q_0 - p^i q_i)^2 + 4\,q_0 q_1 q_2 q_3 + 4\,p^0 p^1 p^2 p^3
\nonumber\\
                              &\?   + 4 (p^1 p^2 q_1 q_2 + p^1 p^3 q_1 q_3 + p^2 p^3 q_2 q_3)\,. 
\end{align}
This frame, despite its non-cubic prepotential, is equivalent to the one in \eqref{eq:STU-def}, but allows for an AdS$_4$ vacuum with a purely electric gauging. In fact, this model arises by consistent truncation of M-theory \cite{Duff:1999gh,Cvetic:1999xp}, so that the corresponding solutions may be oxidized to eleven-dimensional supergravity. 

We finish this subsection by briefly mentioning two simple models that can be obtained by truncations of \eqref{eq:STU-def} and \eqref{eq:STU-root}. In particular, the T$^3$ model is found by setting all three scalars in the STU model to be equal, with the result
\begin{align}\label{eq:T3-model}
F^{\scriptscriptstyle T^3}= &\? \frac{(X^1)^3}{X^0}\,, 
\nn\\
I_4(\Gamma)^{\scriptscriptstyle T^3}= &\? - (q_0 p^0)^2 - 6\? q_0 p^0 q_1 p^1 + 3\? (q_1 p^1)^2 + 4\? q_0 (p^1)^3 - 4\? p^0 (q_1)^3\ .
\end{align}
Similarly, one may obtain the minimally coupled $X^0 X^1$ model
\begin{align}\label{eq:X0X1-model}
F^{X^0 X^1} = i X^0 X^1\ ,
\qquad
I_4(\Gamma)^{X^0 X^1}= (q_0 q_1 + p^0 p^1)^2 \ ,
\end{align}
by identifying $X^2=X^0$ and $X^3=X^1$ in the prepotential \eqref{eq:STU-root} and appropriately rescaling the scalars and charges.

\subsection{BPS equations}
\label{subsec:BPS}

The BPS equations for solutions of abelian gauged $\mathcal{N}=2$ supergravity with a timelike Killing vector were given in \cite{Cacciatori:2008ek, Meessen:2012sr, Chimento:2015rra}. We start from the BPS equations as summarized conveniently given in the latter paper for the timelike class, for which the metric takes the form \eqref{eq:metr-bps}, as appropriate for black hole solutions. 

The BPS equations fix the gauge field strengths $\cF$ in \eqref{eq:F-time-dec} in terms of the scalars, so that the Maxwell equations and Bianchi identities take the form of a Poisson equation on the base metric, as 
\begin{gather}\label{eq:Poiss-Ortin}
d \cF = 
-d\left[ \star d \sv - 2\, \ex^{-4\?U}\,\Iprod{\star \hat{G}}{\qv}\,\qv + \? \ex^{-2\?U}\mathrm{J}\star\hat{G} \right] +d \omega\wedge \hat{G} 
= 0\,,
\end{gather}
where $\mathrm{J}$ is the scalar dependent complex structure in \eqref{CY-hodge}, $\hat{G}$ is the direct product of the vector of gaugings with a one-form\footnote{This is necessarily so for theories without hypermultiplets, while in the more general case $\hat{G}$ is replaced by the hypermultiplet moment maps.}. Introducing the vielbeine $e^x$, with $x$, $y \dotso =1,2,3$, for the three-dimensional base metric $ds^2_3$, $\hat{G}$  must satisfy the equation
\begin{equation}\label{eq:de-bps}
 d e^x - \Iprod{\hat{G}}{\sv} \wedge e^x + \varepsilon^{xyz}\Iprod{\cal A}{\hat{G^y}}\wedge e^z =0 \,,
\end{equation}
where ${\cal A}$ denotes the spatial gauge potentials in \eqref{eq:F-time-dec}. The final BPS equation imposes that the rotation one-form $\omega$ must satisfy
\begin{align}
\star d\omega = &\, \Iprod{d \sv}{\sv} - 2\? \ex^{-4\?U}\,\Iprod{\hat{G}}{\qv}   \label{eq:omega-eqn-gen}
\\
= &\, \ex^{-4\?U}\,\Iprod{\qv}{d \qv + 2 \?\hat{G} }  \label{eq:omega-eqn-gen-2}\,,
\end{align}
where in the second line we re-expressed the first term through the variable $\qv$. The conditions \eqref{eq:Poiss-Ortin}, \eqref{eq:de-bps} and \eqref{eq:omega-eqn-gen} are sufficient to preserve supersymmetry.

The Poisson equation \eqref{eq:Poiss-Ortin} guarantees the local existence of the spatial gauge field strengths, $\cF$, which can be obtained by writing this equation as a total derivative. However, this is subtle in general due to the last term on the RHS, as it can be written as a total derivative in more than one way. In particular, invariance of $\cF$ following from \eqref{eq:Poiss-Ortin} under time reparametrizations, $\omega \rightarrow \omega + d \sigma$, for any function $\sigma$ on the base, requires that $\hat G$ be exact on a simply connected manifold. We write 
\begin{equation}\label{eq:r-def}
 \hat G = G\? d\rho \,,
\end{equation}
for some function $\rho$, to be determined by \eqref{eq:de-bps} once a particular base is chosen. One may then recast \eqref{eq:Poiss-Ortin} as the flow equation
\begin{gather}\label{eq:fl-Ortin}
 \star d \sv - 2\, \ex^{-4\?U}\,\Iprod{\star \hat{G}}{\qv}\,\qv + \? \ex^{-2\?U}\mathrm{J}\star\hat{G} -\rho\? d \omega\?  G  + \cF = 0 \,.
\end{gather}
Here, we note that our choice of total derivative is consistent with the complex anti-selfduality condition \eqref{cmplx-sdual}, where the time components of the gauge fields are given by \eqref{eq:zeta-F} as
\begin{equation}\label{eq:zeta-BPS}
 d\zeta = d\qv - \hat{G}\,.
\end{equation}

The flow equation \eqref{eq:fl-Ortin} can be simplified using the identity \eqref{I4toJ}, repeated here in terms of the variables $\qv$ and $\sv$
\begin{align}  \label{I4toJ-2}
 \frac12\, I^\prime_4(\sv , \sv, \Gamma)  = 2\,\Iprod{\Gamma}{\sv}\,\sv +4\,\ex^{-4\?U}\,\Iprod{\Gamma}{\qv}\,\qv - 2\,\ex^{-2\?U}\mathrm{J}\,\Gamma \,, 
\end{align}
leading to
\begin{align}\label{eq:fl-Ortin-s}
\star d \sv + \Iprod{\star \hat{G}}{\sv}\,\sv 
-\frac14\? I^\prime_4(\sv , \sv , \star\hat{G}) &\, -\rho\? d \omega\? G + \cF = 0\,.
\end{align}
In this form, the flow equation is algebraic in terms of the combination $\sv$, allowing for simpler manipulation using techniques similar to \cite{Katmadas:2014faa, Halmagyi:2014qza}, as will be seen below.

In all the cases we consider we are interested in black holes corresponding to a radial flow from the asymptotic region to the near-horizon geometry. We can therefore already specify an ansatz for the three-dimensional base space as a product of the radial direction with a 2d surface $\Sigma$, as
\begin{equation}\label{eq:3d-metr}
d s_3^2 = dr^2 + \ex^{2 \psi (r)}\? ds^2_{\Sigma}\ .
\end{equation}
We can then decompose the vielbein in terms of the veilbein $\hat{e}^{a}$ on the surface $\Sigma$
\begin{align}
 e^1 = dr\,, \qquad  e^a = \ex^\psi\ \hat{e}^{a}\,,
\end{align} 
where the indices $a, b \dotso=2,3$.
For this basis it turns out the function $\rho$ associated with the one-form $\hat{G}$ becomes the radial coordinate, such that
\begin{equation}
	\hat{G} = G\ dr\ .
\end{equation}
With this ansatz we find further simplifications in \eqref{eq:de-bps} and \eqref{eq:fl-Ortin-s} so that we can finally present the full set of BPS equations in the compact form
\begin{subequations}
\label{eq:final}
 \begin{empheq}[box=\fbox]{align}
 \label{eq:final-psi}
\psi' =&\, \Iprod{G}{\sv}\ , \rule[.5cm]{0cm}{0cm}  \\
 \label{eq:final-Sig}
\hat{\omega}^{ab} = &\, \varepsilon^{ab}\?\Iprod{G}{{\cA}}\ , 
\\
 \label{eq:final-omega}
  \star d\omega = &\, \Iprod{d \sv}{\sv} + \Iprod{G}{I^\prime_4(\sv)}\ d r\ , 
\\
 \label{eq:final-flow}
\quad \ex^{-\psi}\ d (\ex^\psi \sv) =&\,
\frac14\? I^\prime_4(\sv , \sv , G)\  d r  + G\ r\ \star d \omega\? - \star \cF\ , \quad {\rule[-.5cm]{0cm}{0cm}}
\end{empheq}
\end{subequations}
where $\varepsilon^{ab} = \varepsilon^{1ab}$ and $\hat{\omega}^{ab}$ denote the antisymmetric symbol and the spin connection on the Riemann surface $\Sigma$, respectively.
In order to obtain black hole solutions, one must solve \eqref{eq:final} for a given electromagnetic charge vector $\Gamma = \{p^I, q_I\}$, defined as
\begin{equation}\label{eq:charge-def}
	\Gamma \equiv \frac{1}{\Vol_\Sigma} \int_{\Sigma} \cF\ ,
\end{equation}
along with boundary conditions for the scalars that will be spelled out in the following section. Note that the requirement \eqref{eq:charge-def} fixes the spatial field strengths $\cF$ completely in the static case in general, as well as in a class of rotating asymptotically flat black holes \cite{Bossard:2012xsa}, but not for rotating black holes in gauged supergravity, as will be seen in due course.

\section{Warm-up: asymptotics and static black holes}
\label{sec:warm-up}

Before proceeding with the analysis of the BPS equations for rotating black holes, we pause to present some known properties of the solutions to \eqref{eq:final}, in order to gain some extra intuition and understanding of the real formulation of supergravity. We focus on the possible asymptotic structure of the solutions depending on the gauging, listing known examples for each case, and then give a short review of the static asymptotically AdS$_4$ black hole solutions.

\subsection{Asymptotic structure and string theory embeddings}
\label{sec:as-vac}
We first consider the asymptotic behaviour of solutions, depending on the model and on the choice of gauging. As it turns out, for scalar manifolds specified by the cubic prepotential and in particular for the STU model, one can asymptotically realize various possibilities, including AdS$_4$, Minkowski space, as well as hyperscaling violating Lifshitz (hvLif) spacetimes\footnote{For these latter hyperscaling violating geometries the potential is not stabilized at infinity: the scalar fields do not reach asymptotically a constant value. With a slight abuse of terminology we will refer to them as "runaway vacua".}, depending on the choice of an explicit gauging vector, $G$. For each of these vacua and types of gauging vectors we then give examples of models with a string/M-theory origin.

We consider the following expansion in the asymptotic region, around $r\rightarrow \infty$, parametrized by a constant $\lambda$ and a constant symplectic vector $\cA_2$, as
\begin{equation}\label{eq:asympt-exp}
 \ex^{2\?\psi}\?\sv = \lambda\, I^\prime_4(G)\, r^3 + \cA_2\, r^2 + \cO(r) \,, 
\end{equation}
from which the function $\ex^\psi$ is determined from \eqref{eq:final-psi} as
\begin{equation}\label{eq:asympt-exp-2}
 \ex^{2\?\psi} = 2\?\lambda\, I_4(G)\, r^4 + \frac23\, \Iprod{G}{\cA_2}\, r^3 + \cO(r^2) \,. 
\end{equation}
One can easily verify that \eqref{eq:asympt-exp}-\eqref{eq:asympt-exp-2} solve the flow equation \eqref{eq:final-flow} up to a first order expansion asymptotically, for any $\lambda$ and any $\cA_2$ that satisfies 
\begin{equation}
 \Iprod{\cA_2}{I^\prime_4(G)} = 0\,.
\end{equation}
The asymptotic expansion for the combination in \eqref{eq:asympt-exp} specifies the relevant boundary conditions for all solutions considered in this paper, depending on the choice for the gauging and the properties of $\cA_2$. We distinguish four cases depending on the properties of the vector $G$:
\begin{itemize}
\item AdS$_4$ vacuum: \quad $I_4 (G) \neq 0$, \quad $G$ is {\it generic} (rank-4),
\item hvLif (class I): \quad $I_4 (G) = 0, \,\, I_4'(G) \neq 0$, \quad $G$ is {\it restricted} (rank-3),
\item hvLif (class II): \quad $I_4 (G) = I_4'(G) = 0, \,\, I_4''(G) \neq 0$, \quad $G$ is {\it small} (rank-2),
\item $\mathbb{R}^{1,3}$ vacuum: \quad $\frac14\,I_4 (G, G, \Gamma_1, \Gamma_2) = - \Iprod{G}{\Gamma_1}\,\Iprod{G}{\Gamma_2}$, \quad $G$ is {\it very small} (rank-1).
\end{itemize}
To make things explicit, let us consider the case of the STU model, for which a gauging vector with components
\begin{equation}
 G = \{ g^0,g^1,g^2,g^3,g_0,g_1,g_2,g_3 \}^{\rm T} \,,
\end{equation}
leads to the following $I_4(G)$ (cf. \eqref{I4-ch-stu})
\begin{align}
I_4(G)^{\scriptscriptstyle STU} = &\? - (\sum_I g_I g^I)^2 + 4\,g_0 g^1 g^2 g^3- 4\,g^0 g_1 g_2 g_3 
\nonumber\\
                              &\?    + 4 (g^1 g^2 g_1 g_2 + g^1 g^3 g_1 g_3 + g^2 g^3 g_2 g_3)\,.
\end{align}
Depending on the choice of nonvanishing components of $G$ we can have all of the options mentioned above. 
We proceed to discuss the properties of the asymptotic geometries of each class that are important in later sections, providing string theory embeddings from reductions of higher dimensional supergravity.

\paragraph{AdS$_4$} 
We start with the generic case, where the gauging vector $G$ is rank-4, so that the constant $\lambda$ is fixed in terms of the AdS$_4$ radius as $\lambda = \frac12\?I_4(G)^{-1/2}$, and \eqref{eq:asympt-exp}-\eqref{eq:asympt-exp-2} become 
\begin{align}\label{eq:as-sol}
 \ex^{2\?\psi}\?\sv =&\? \frac1{2\?\sqrt{I_4(G)}}\, I^\prime_4(G)\, r^3 + {\cal A}_2\, r^2 + \cO(r) \,, 
 \nonumber\\
 \ex^{2\?\psi} =&\? \sqrt{I_4(G)}\, r^4 + \frac23\,\Iprod{G}{{\cal A}_2}\, r^3 + \cO(r^2) \,, 
\end{align}
It follows that the four dimensional metric \eqref{eq:metr-bps} for the base \eqref{eq:3d-metr} becomes asymptotically
\begin{equation}
 ds^2_{AdS_4} = - \sqrt{I_4(G)}\,r^2\?dt^2 +\frac{1}{\sqrt{I_4(G)}}\, \left( \frac{dr^2}{r^2} + 2\,I_4(G)\?r^2\? ds^2_{\Sigma} \right)\,,
\end{equation}
which is the standard metric on AdS$_4$ up to rescaling of the time and radial coordinates.

An example generic vector is given by 
\begin{equation}
G = \{ g^0,0,0,0,0,g_1,g_2,g_3 \}^{\rm T}\quad \Rightarrow\quad I_4(G) = - 4\,g^0 g_1 g_2 g_3 \,,
\end{equation}
for the well known case of the gauged $STU$ model in 4d. The choice $-g^0 = g_1 = g_2 = g_3 = 1$, after symplectically rotating as in \eqref{eq:STU-sym-rot}, corresponds to the STU model in the frame \eqref{eq:STU-root}, which arises from maximal 4d gauged supergravity and thus from a direct reduction of 11d supergravity on S$^7$ \cite{Cvetic:1999xp}.

\paragraph{hvLif:}
A less restricted class of asymptotics is given by the hyperscaling violating Lifshitz (hvLif) runaway vacua, a two-parameter family of asymptotic geometries that includes AdS space as a special case. For the general definition and main properties of hvLif solutions we refer to \cite{Perlmutter:2012he}. Here, we use the parametrization of the hvLif runaway vacua in terms of two constants, the dynamical exponent $z$ and the hyperscaling violation exponent $\theta$, given by
\begin{equation}
 ds^2_{hvLif} = r^{-\theta}\left( - r^{2\?z}\, dt^2 + \frac{dr^2}{r^2} + r^2\? ds^2_{\Sigma} \right)\,,
\end{equation}
which reduces to AdS$_4$ for $z = 1, \theta = 0$. We divided the hvLif solutions in class I and II depending on the rank of the gauging vector $G$, but in fact there are various possibilities for the values of $z$ and $\theta$ in each case, arising by imposing constraints on the subleading terms specified by the vector ${\cal A}_2$. Instead of giving a general discussion, we restrict ourselves here to the explicit examples of interest for us.

Within the rank-3 class for $G$ (class I), the metric functions turn out to scale as $\ex^{U} \sim r^{1/2}$, $\ex^{\psi} \sim r^{3/2}$ in the general case, leading to the exponents $z=0$ and $\theta=-2$.
However, in this paper we are interested in solutions that can be lifted to asymptotically AdS$_5$ geometries. This is achieved by imposing the constraint
\begin{equation}
 I_4(I^\prime_4(G),I^\prime_4(G),\cA_2,\cA_2) =0\,,
\end{equation}
on the subleading vector $\cA_2$.
In this special case, the expansion in \eqref{eq:asympt-exp}-\eqref{eq:asympt-exp-2} describes a behaviour where $\ex^{2\?U} \sim \ex^{\psi} \sim r^{3/2}$.
The asymptotic four dimensional metric then takes the form
\begin{equation}
 ds^2_{hvLif_I} = - r^{3/2}\,dt^2 + \frac{dr^2}{r^{3/2}} + r^{3/2}\? ds^2_{\Sigma} \,,
\end{equation}
which, after redefining the radial variable as $r=\rho^2$ and rescaling the other coordinates by a constant factor, takes the form
\begin{equation}\label{eq:metr-rank-3}
 ds^2_{hvLif_I} = \rho\left( - \rho^2\, dt^2 + \frac{d\rho^2}{\rho^2} + \rho^2\? ds^2_{\Sigma} \right)\,.
\end{equation}
The metric \eqref{eq:metr-rank-3} is of the hvLif type with exponents $\theta=-1$ and $z=1$, in the notation of \cite{Perlmutter:2012he}.

A relevant example within the STU model that we use in the following is given by 
\begin{equation}\label{eq:rank3-vec}
G = \{ 0,0,0,0,g_0,g_1,g_2,g_3 \}\quad \Rightarrow\quad 
I_4(G) = 0,\quad (I_4')_0 (G) = \frac{\partial I_4 (G)}{\partial g^0} = -4\, g_1 g_2 g_3\,,
\end{equation}
This possibility, with $g_1 = g_2 = g_3 = \frac12$ arises from a dimensional reduction of five-dimensional $STU$ gauged supergravity which is a further truncation of maximal 5d supergravity. Solutions of this theory have interpretation as asymptotically AdS$_5 \times $S$^5$ solutions in type IIB supergravity. The four-dimensional hvLif$_I$ runaway vacuum indeed corresponds to the dimensional reduction of AdS$_5$ as discussed in \cite{Hristov:2014eza,Hosseini:2017mds}. We will not be further concerned here with hvLif solutions of class II, coming from rank-2 gaugings.

\paragraph{Minkowski} For the case of $G$ being very small, the potential \eqref{eq:3d-pot} vanishes upon using \eqref{eq:Hab-tensor}, so that the bosonic Lagrangian is identical to that of ungauged supergravity, even though $G$ still appears in the gravitino coupling \eqref{eq:gravitino}. Since Minkowski space is not a supersymmetric vacuum in abelian gauged supergravity, asymptotically flat solutions are necessarily non-BPS. Nevertheless, one may still construct such black hole solutions starting from supersymmetric attractors of the type described in \cite{Hristov:2012nu}, so we briefly record this case as well.

An example very small vector is given by 
\begin{equation}\label{eq:rank1-vec}
G = \{ 0,0,0,0,g_0,0,0,0 \}^{\rm T}\quad \Rightarrow\quad I_4(G) = 0,\quad I_4'(G) = 0\,,
\end{equation}
for any cubic symmetric model.
This possibility arises explicitly from a Sherk-Schwarz reduction of five-dimensional ungauged supergravity with a symmetric scalar manifold, as explained in \cite{Looyestijn:2010pb}. In turn this means one can view solutions of this theory as solutions to 11-dimensional supergravity on appropriate CY$_3\times S^1$ compactifications with a twist along the circle.

\subsection{Review of static AdS\texorpdfstring{$_4$}{4} black holes}
\label{sec:static-bh}
As an illustration of the real formulation for the BPS equations, we present here a short summary of the known static BPS black hole solutions in gauged supergravity.

It is instructive to first consider \eqref{eq:final-psi}. We already used the ansatz in \eqref{eq:3d-metr} where the function $\psi$ only depends on the radial coordinate, therefore we can choose an ansatz for the symplectic variable $\sv$ in \eqref{eq:dual-coo} in the following way
\begin{equation}\label{eq:section-ansa}
	\sv = \ex^{- \psi}\ {\cal H} (r) + f (r, \Sigma)\ G\ .
\end{equation}
Here, $\cH (r)$ is a radially dependent symplectic vector and $f(r, \Sigma)$ is an arbitrary function on the base \eqref{eq:3d-metr}, i.e. that can depend also on the coordinates on the Riemann surface $\Sigma_g$. Note that we have chosen the second term in \eqref{eq:section-ansa} proportional to $G$, so that it drops out of \eqref{eq:final-psi}, due to the antisymmetry of the symplectic inner product, $\Iprod{G}{G} = 0$, but in principle any vector that is mutually local with $G$ could be used to the same effect, so that we arrive at the equation
\begin{equation}
	(\ex^\psi)' = \Iprod{G}{{\cal H}}\ .
\end{equation}
In the static case it can further be shown that $f(r, \Sigma) = 0$ but we make use of this freedom in the equation in the construction of the rotating solutions in the following sections. 

We now restrict to the static case, for which the general black hole solution is known \cite{Halmagyi:2014qza}. However, for illustrational purposes, we further restrict to the simpler class of solutions appearing in \cite{Cacciatori:2009iz,Dall'Agata:2010gj,Hristov:2010ri, Katmadas:2014faa} for which $\cH$ is linear in the radial variable\footnote{We remark on the most general case later in this subsection, around \eqref{eq:gen-stat}.}
\begin{equation}\label{eq:stat-ans}
	\cH(r) = {\cal H}_0 + {\cal H}_{\infty}\ r\ ,
\end{equation}
for some constant symplectic vectors ${\cal H}_0$ and ${\cal H}_\infty$ that are to be fixed by the BPS equations. The names are appropriately chosen as ${\cal H}_0$ gives the near-horizon ($r \rightarrow 0$) value of the scalars, while ${\cal H}_\infty$ fixes the scalars asymptotically ($r \rightarrow \infty$). From the discussion of the asymptotics in the previous subsection, we can already see from \eqref{eq:asympt-exp} that
\begin{equation}\label{eq:H-infty}
	{\cal H}_\infty = \lambda I_4' (G)\ ,
\end{equation}
for a constant $\lambda$ that is usually chosen by convention depending on the type of asymptotics. The explicit form will help us simplify substantially the remaining equations using the various identities in App. \ref{app:I4}. Notice that we can already give the general solution for the metric function $\psi (r)$,
\begin{equation}\label{eq:psi-static}
	\ex^\psi = \Iprod{G}{{\cal H}_0}\ r + 2 \lambda\ I_4 (G)\ r^2\ ,
\end{equation}
where $\lambda = \frac12\?I_4(G)^{-3/4}$ to match the AdS$_4$ asymptotics in \eqref{eq:as-sol} and we did not allow for an integration constant that can always be shifted away by a coordinate redefinition.

We then turn to Eq.\ \eqref{eq:final-omega}, whose left hand side vanishes for static solutions, since the rotation one-form $\omega$ vanishes in this case. After a bit of rewriting on the right hand side, one is left with the following constraint:
\begin{equation}
	2\ \ex^\psi\ \Iprod{{\cal H}_0}{{\cal H}_\infty} = \Iprod{G}{I_4' ({\cal H}_0 +{\cal H}_\infty\ r)}\ ,
\end{equation}
which we can further expand in powers of $r$. The left hand side only has a linear and quadratic piece in $r$ by virtue of \eqref{eq:psi-static}, while the right hand side also has a constant and a cubic piece in $r$ since $I_4'$ is homogeneous of degree 3. This in principle amounts to four equations, but upon the explicit use of \eqref{eq:H-infty} together with the identities \eqref{eq:quint}-\eqref{eq:proj}, it turns out three of these equations are already satisfied. One constraint for the vector ${\cal H}_0$ remains, 
\begin{equation}
	I_4 ({\cal H}_0, {\cal H}_0, {\cal H}_0, G) = 0 \quad \Rightarrow \quad \Re(\ex^{-\im\?\alpha}Z(G))\bigr|_{\text hor}=0\ ,
\end{equation}
where in the second equation we rewrote the constraint in the complex basis, recognized as one of the BPS equations at the attractor.

Proceeding with \eqref{eq:final-flow}, we note that due to the symmetries we can set the gauge field strengths proportional to the volume form of the Riemann surface, ${\cal F} = \Gamma\ {\rm vol}_{\Sigma}$. We therefore find
\begin{equation}
	\ex^\psi\ {\cal H}_\infty = \frac14 I_4' ({\cal H}_0 + {\cal H}_\infty\ r, {\cal H}_0 + {\cal H}_\infty\ r, G) - \Gamma\ ,
\end{equation}
leading to three vector equations once expanded in powers of the radial variable. In particular, the quadratic terms in $r$ cancel upon inserting the solution for ${\cal H}_\infty$ in \eqref{eq:H-infty}. Using the identity \eqref{eq:proj}, we then remain with the condition
\begin{equation}\label{eq:static-constr}
 \Iprod{{\cal H}_0}{{\cal H}_\infty} = 0\,,
\end{equation}
from the terms linear in $r$ and with
 \begin{empheq}[box=\fbox]{align}\label{eq:H-0}
	\Gamma = \frac14 I_4' ({\cal H}_0, {\cal H}_0, G)\ ,
\end{empheq}
at the horizon at $r=0$. The latter is an equation for a full symplectic vector and is therefore enough to completely solve for the near-horizon symplectic vector ${\cal H}_0$ in terms of the gauging vector $G$ and the charge vector $\Gamma$. Eventually the constraint \eqref{eq:static-constr} can be transformed into a restriction on the allowed electromagnetic charges allowed by the gauging,
 \begin{empheq}[box=\fbox]{align}\label{eq:final-constr}
	I_4 (\Gamma, \Gamma, \Gamma, G)  = I_4 (G, G, G, \Gamma)  = 0\ .
\end{empheq}
We have therefore written down the solution for a static black hole subject to a particular constraint on the charges. In order to find the most general solution we should have started with a slightly more general ansatz for the section
\begin{equation}\label{eq:gen-stat}
	\sv = e^{-2 \psi} \left( {\cal A}_1\ r + {\cal A}_2\ r^2 + {\cal A}_3\ r^3 \right)\ ,
\end{equation}
for three constant symplectic vectors ${\cal A}_{1,2,3 }$ determined by the equations. In particular the AdS$_4$ asymptotics fix $\cA_3 = \frac12\?I_4(G)^{-1/2}\,I_4' (G)$. The complete solution can be found in a similar fashion, as can be seen explicitly in \cite{Halmagyi:2014qza}, but is slightly more complicated due to the more general ansatz. In this case the constraint \eqref{eq:final-constr} is relaxed, thus allowing for one extra free parameter in the charge vector \cite{Halmagyi:2014qza}.

Finally let us remark on the last equation still to be solved, \eqref{eq:final-Sig}. Upon taking an exterior derivative we can then relate the gauge field strengths to the Ricci curvature of the 2d surface,
\begin{equation}\label{eq:charge-constr}
	\Iprod{G}{{\cal F}} = \epsilon_{ab} \hat{{\cal R}}^{ab}\ ,
\end{equation}
where $\hat{{\cal R}}^{ab}$ is the Ricci form on $\Sigma$.
In the static case all isometries of the 2d surface $\Sigma$ are preserved and one can further show that near the horizon the 2d metric must be the constant curvature metric, which we can also take as a solution along the full black hole flow (see \cite{Anderson:2011cz} for a more careful analysis). This leaves three possibilities, namely the constant curvature metrics on $S^2$ or H$_2$ and the flat metric on ${\mathbb R}^2$. In order to have a compact horizon we can further quotient the non-compact spaces ${\mathbb R}^2$ and H$_2$ to arrive at a Riemann surface $\Sigma_{\fg}$ of arbitrary genus. Explicitly, one can use any of the following metrics
\begin{equation}
ds^2_{\Sigma_\fg} = d\theta^2 + f_\kappa^2(\theta)\, d\phi^2 \;,\quad 
f_\kappa(\theta) = \left\{ 
\begin{array}{ll} \sin\theta & \kappa = 1 \\ 1 & \kappa=0 \\ \sinh\theta & \kappa=-1\,, \end{array} \right.
\end{equation}
where $\kappa = 1$ for $S^2$, $\kappa = 0$ for $T^2$, and $\kappa = -1$ for $\Sigma_\fg$ with $\fg>1$.
The scalar curvature in each case is $2\?\kappa$ and the volume is
\begin{equation}
\Vol(\Sigma_\fg) = 2\? \pi \?\eta \;, \quad 
\eta = \left\{ \begin{array}{ll} 2\? |\fg-1| &\text{for } \fg \neq 1 \\ 1 &\text{for } \fg = 1 \;. \end{array} \right.
\end{equation}
The gauge field strengths are then given by 
\begin{equation}
	 {\cal F} = \Gamma\ f_\kappa(\theta)\ d\theta \wedge d\phi\ ,
\end{equation}
and we finally arrive (using \eqref{eq:charge-def} and integrating \eqref{eq:charge-constr}) at one final constraint on the charges,
 \begin{empheq}[box=\fbox]{align}\label{eq:final-charge-constr}
	\Iprod{G}{\Gamma} = -\kappa\ .
\end{empheq}
Summarizing, in order to obtain a static BPS solution in the class described by \eqref{eq:stat-ans}, one has to solve \eqref{eq:H-0}, subject to the constraints \eqref{eq:final-constr} and \eqref{eq:final-charge-constr}. In the next sections, we will extend these solutions to the rotating case by finding solutions of the form \eqref{eq:section-ansa}.

\section{Rotating near-horizon geometry}
\label{sec:nhg}

\subsection{Solving the BPS conditions}
\label{subsec:nhg-BPS}

Restricting to an attractor solution, which we expect to be topologically AdS$_2\times \Sigma$, we impose the appropriate scaling with respect to the radial coordinate for all the relevant fields. In particular, we take the function $\psi(r)$ in \eqref{eq:3d-metr} as
\begin{equation}\label{eq:psi-atrr}
 \ex^\psi = \vvv \, r \,,
\end{equation}
 with $\vvv$ a constant that physically gives the ratio between the scales of $\Sigma$ and AdS$_2$ on the horizon, so that the three-dimensional base metric becomes
\begin{equation}\label{eq:base-bps-attr}
 ds^2_3 = dr^2 + \vvv^2\, r^2 \? ds^2_{\Sigma}\,.
\end{equation}
The conical structure of this ansatz implies the scaling behaviour
\begin{equation}\label{eq:eU-attr}
 \ex^{-2\?U} = \frac{1}{r^2}\? \ex^{-2\? \u}\,, \qquad \omega = \frac{1}{r}\?\omega_0 \,,
\end{equation}
for the remaining objects in the 4d metric, where $\u$ and $\omega_0$ are a function and a one-form that depend only on the coordinates on $\Sigma$. The total metric \eqref{eq:metr-bps} thus takes the expected form
\begin{equation}\label{eq:metr-bps-atrr}
ds^2_4 = -\ex^{2\?\u} \left(r \? dt +  \omega_0 \right)^2 + \ex^{-2\?\u}\?\left( \frac{1}{r^2} \? dr^2 + \vvv^2 \, ds^2_{\Sigma} \right)  \,.
\end{equation}
The requirement \eqref{eq:eU-attr} implies that the variables $\zeta$, $\qv$ and $\sv$ behave as
\begin{equation}\label{eq:scal-fields}
 \zeta = r\? \zeta_0 \,, \qquad  \qv = r\? \qv_0 \,, \qquad  \sv = \frac1r\? \sv_0 \,, 
\end{equation}
where $\zeta_0$, $\qv_0$ and $\sv_0$ are symplectic vectors that depend only on the coordinates on $\Sigma$ and are such that
\begin{equation}
 \ex^{4\?\u} = I_4(\qv_0) = I_4(\sv_0)^{-1} \,.
\end{equation}
Finally, we note that with the choice \eqref{eq:psi-atrr}, the condition \eqref{eq:final-psi} becomes
\begin{equation}\label{eq:ImG-atrr}
  \Iprod{G}{\sv_0} = 1 \,.
\end{equation}

Turning to the flow equation in \eqref{eq:final-flow}, the scaling behaviour \eqref{eq:psi-atrr}-\eqref{eq:scal-fields} leads to 
\begin{align}\label{eq:fl-Ortin-f}
\frac1r\? \star d \sv_0  -\frac1{4\?r^2}\? \ex^{-2\?\u}I^\prime_4(\sv_0 , \sv_0 , G)\? \star dr - r\? d\left( \frac{1}{r}\?\omega_0 \right) \?G + \cF = 0\,.
\end{align}
By the assumption that the scalar fields and $\ex^\u$ do not depend on the radial coordinate near the horizon, this flow equation breaks up in components as 
\begin{gather}
\star d \sv_0  = \omega_0 \wedge dr\, G \,, \label{eq:d-scalars}
\\
\cF = \frac1{4\?r^2}\? \ex^{-2\?\u}I^\prime_4( \sv_0 , \sv_0 , G )\? \star dr + d\omega_0 \?G\,. \label{eq:attractor}
\end{gather}
Here, the first equation determines the dependence of the scalars along $\Sigma$, while the second fixes their constant parts in terms of the charges, directly generalizing the corresponding static attractor equation.
The final condition to reduce on $\Sigma$ is \eqref{eq:final-omega}. To this end, note that taking the inner product of \eqref{eq:d-scalars} with $\sv_0$ and using \eqref{eq:ImG-atrr}, one finds
\begin{equation}
 \Iprod{d \sv_0}{\sv_0} = \star\left( \omega_0\?\wedge dr \right) \,,
\end{equation}
so that \eqref{eq:final-omega} becomes
\begin{align}\label{eq:omega-eqn}
\star d \omega_0
= &\? \frac1{r^2}\? \Iprod{G}{I^\prime_4(\sv_0)} \? dr\,.
\end{align}
Acting with a derivative on \eqref{eq:omega-eqn} and using \eqref{eq:d-scalars}, we find
\begin{equation}
 d\star d \omega_0 = \frac1{2\? r^2}\?I_4\left( \sv_0 , \sv_0 , G, G \right)\, \star\left( \omega_0\wedge dr \right)\wedge dr\,.
\end{equation}
The expression in the RHS is recognized as the contraction of \eqref{eq:attractor} with $G$, which is in turn fixed by \eqref{eq:final-Sig}, leading to
\begin{equation}\label{eq:Lapl-omega}
 d\star_\scr2 d \omega_0  = \mathrm{R}^\scr2 \? \star_\scr2 \omega_0 \,,
\end{equation}
where $\star_\scr2$ and $ \mathrm{R}^\scr2$ are the Hodge star and Ricci scalar on $\Sigma$. We therefore conclude that $\omega_0$ must be the one-form dual to a Killing vector field $\tilde\omega_0$ on $\Sigma$. Unlike the static case we discussed above, this in turn means that rotating black holes can only have spherical topology if we insist on compactness. Rotating black holes with non-spherical topology have non-compact horizons, i.e. cylindrical or hyperbolic ones, see  \cite{Caldarelli:1998hg}

With this knowledge we can now be more specific and write down the general form of the metric on the 2d space $\Sigma$. As it turns out, this cannot be of the constant curvature type, as it will become clear by the following analysis that such a choice would be inconsistent for a gauging allowing for an AdS$_4$ vacuum. We therefore parametrize the metric on $\Sigma$ in terms of a single function $\Delta(\theta)$, as
\begin{equation}\label{eq:base-bps-sph}
 ds^2_{\Sigma} =  \frac{d\theta^2}{\Delta(\theta)} + \?\Delta(\theta)\? f_\kappa^2(\theta)\, d\phi^2 \, ,\quad 
f_\kappa(\theta) = \left\{ 
\begin{array}{ll} \sin\theta & \kappa = 1 \\ 1 & \kappa=0 \\ \sinh\theta & \kappa=-1 \end{array} \right.
\end{equation}
where $\kappa = 1$ for the spherical case, $\kappa = 0$ for the cylindrical, and $\kappa = -1$ for the hyperbolic. 
Here, $\theta$, $\phi$ are coordinates along the surface $\Sigma$, and $\Delta(\theta)$ is a function of $\theta$. The direction $\phi$ corresponds to a compact isometry in all three cases. 
We also record the Ricci scalar corresponding to the metric \eqref{eq:base-bps-sph}, given by
\begin{equation}\label{eq:Ric-Sigma}
\mathrm{R}^\scr2 = -\frac{1}{f_\kappa}\?\partial_\theta\left( \frac{1}{f_\kappa}\?\partial_\theta \left( \Delta(\theta)\? f^2_\kappa\! (\theta) \right) \right)  \,.
\end{equation}

One can now insert \eqref{eq:Ric-Sigma} into \eqref{eq:Lapl-omega}, which can be readily solved for $\omega_0$ as
\begin{equation}\label{eq:omega-sph}
 \omega_0 = -\frac{\Jpar}{\vvv}\?\Delta(\theta)\? f^2_\kappa\! (\theta)\,d\phi  \,,
\end{equation}
where $\Jpar$ is a constant and the factor of $\vvv$ was added for later convenience. Note that the dual vector to \eqref{eq:omega-sph} with respect to the metric \eqref{eq:base-bps-sph} is simply $\tilde\omega_0 =-\frac{\Jpar}{\vvv}\?\frac{\partial}{\partial\? \phi}$, which is manifestly Killing and corresponds to a compact $U(1)$ isometry. With this expression for $\omega_0$, we can explicitly compute 
\begin{equation}
 \star\left( \omega_0\wedge dr \right) = \frac{\Jpar}{\vvv}\? d(F_\kappa)\,,
\end{equation}
where $\partial_\theta F_\kappa \equiv - f_\kappa$, so that \eqref{eq:d-scalars} can be solved as
\begin{equation}\label{eq:ImO-attr}
 \vvv\?\sv_0  = e^{\psi} \sv = {\cal H}_0 + \Jpar\? F_\kappa\? G\,,
\end{equation}
for a constant symplectic vector, ${\cal H}_0$. Consistency with \eqref{eq:ImG-atrr} requires that
\begin{equation} \label{eq:vvv}
	\vvv = \Iprod{G}{{\cal H}_0}\ .
\end{equation}

Having worked out the BPS conditions for the different choices of $\Sigma$, from here on we specialize to the spherical case in order to be explicit. We come back to present the general formulas for other cases at the end of this section.
Using the functions $f_\kappa = \sin \theta, F_\kappa = \cos \theta$ in \eqref{eq:attractor}, we obtain the explicit form of the gauge field strengths for a spherical horizon, as
\begin{equation}\label{eq:F-attr}
 \cF = \cB\? \sin \theta\ d\theta\wedge d\phi + d\omega_0 \?G \,,
\end{equation}
where the $\cB$ are given by
\begin{equation}\label{eq:attr-fin0}
 \cB = \frac{1}{4}\? I^\prime_4\left({\cal H}_0 + \Jpar\? \cos \theta\ G , {\cal H}_0 + \Jpar\? \cos \theta\ G, G \right) \,,
\end{equation}
and we write the term along the gauging separately, as it does not contribute to the charge and ultimately to the attractor equation.
The charge vector is obtained through its standard definition \eqref{eq:charge-def}, as
\begin{equation}
 \Gamma = \frac{1}{\Vol_\Sigma}\?\int_{\Sigma} \cF = \frac{1}{\Vol_\Sigma}\?\int_{\Sigma} \cB\? \sin \theta\ d\theta\wedge d\phi \,.
\end{equation}
Using \eqref{eq:attr-fin0} explicitly, we find 
 \begin{empheq}[box=\fbox]{align}\label{eq:attr-fin}
  \Gamma = \frac{1}{4}\? I^\prime_4\left({\cal H}_0, {\cal H}_0, G \right) + \frac{1}{2}\? \Jpar^2\? I^\prime_4\left( G \right)\,,
\end{empheq}
which is the final attractor equation to be solved for ${\cal H}_0$, the constant part of the scalars. It generalizes the static attractor equation \eqref{eq:H-0} and can be explicitly solved in a given model defined by a prepotential (which in turn defines the quartic invariant $I_4$) and a gauging vector $G$.

The final object is the function $\Delta(\theta)$, which appears in the attractor metric \eqref{eq:metr-bps-atrr} both in the base space metric and in $\omega$. This is obtained from \eqref{eq:omega-eqn} upon using the expression \eqref{eq:ImO-attr} for the scalar section, leading to
\begin{align} \label{eq:omega-attr}
\omega_0 = & - \frac{\Jpar}{\vvv}\? \left( 1  - \Iprod{{\cal H}_0}{I^\prime_4(G)}  \? \Jpar \? \cos \theta\ + I_4(G)\? \Jpar^2 \sin^2 \theta\ \right)\? \sin^2 \theta\ \?d\phi
\nonumber\\
&\? -\frac1{\vvv}\?\left( \Iprod{G}{I^\prime_4({\cal H}_0)}  + \Iprod{\cH_0}{I^\prime_4(G)}  \?  \Jpar^2 \right)\? \cos \theta\ d\phi \,,
\end{align}
where we used the inner product of \eqref{eq:attr-fin} with $G$ to rearrange terms, together with the final constraint from \eqref{eq:final-Sig} 
 \begin{empheq}[box=\fbox]{align}
 \label{eq:final-sph}
\Iprod{G}{\Gamma} = -1\ .
\end{empheq}
The second line in \eqref{eq:omega-attr} corresponds to a NUT charge, which must vanish for a regular solution\footnote{The presence of nonzero NUT charge requires a compact time to avoid Misner strings. Therefore the solution would have closed timelike curves.}. We thus arrive at
\begin{equation}\label{eq:Delta-sol}
\Delta(\theta) =  1  - \Iprod{\cH_0}{I^\prime_4(G)}  \? \Jpar \? \cos \theta\ + I_4(G)\? \Jpar^2 \sin^2 \theta\ \,,
\end{equation}
along with 
\begin{equation}\label{eq:NUT-ch}
N =\frac1{2\vvv} \left( \Iprod{G}{I^\prime_4(\cH_0)}  +  \Jpar^2\ \Iprod{\cH_0}{I^\prime_4(G)} \right)\ ,
\end{equation}
which we set to zero here for a regular rotating black hole\footnote{See \cite{Erbin:2015gha} for solutions with NUT charge in gauged supergravity.}.

If we choose the internal space to be non-compact, i.e.\ the cylindrical and hyperbolic rotating black holes, most of the above formulas generalize easily by inserting $\kappa$ where appropriate. In particular we have $\Iprod{G}{\Gamma} = - \kappa$
and we again need to solve the same main equation \eqref{eq:attr-fin} in order to find the full black hole metric and the scalars. In the explicit examples below we concentrate on the spherical solutions, but also give one example with non-compact horizon in order to also relate with previous literature.

\subsection{Summary of BPS attractors}
\label{sec:summ-attr}

Given the above results, we now summarize the structure of rotating BPS attractors. The metric is as in \eqref{eq:metr-bps-atrr}, with the metric along the Riemann surface as in \eqref{eq:base-bps-sph}, where the function $\Delta$ is given by \eqref{eq:Delta-sol}. The rotation one-form $\omega_0$ is given by \eqref{eq:omega-sph}, while the NUT charge is given by \eqref{eq:NUT-ch}. The remaining fields are given in terms of the constant vector $\cH_0$, which is determined as the solution to \eqref{eq:attr-fin}. In particular, the constants $\ex^\u$ and $\vvv$, along with the scalar fields are given by \eqref{eq:ImO-attr} and \eqref{eq:vvv}. More explicitly, the combination $\vvv\?\ex^{-\u}$ is computed through
\begin{equation}\label{eq:e4u-atrr}
 \vvv^4\? \ex^{-4\?\u} = I_4(\cH_0 + \Jpar\? \cos \theta\ G) = \cW + \Jpar^2\? \Delta(\theta)\?  \sin^2 \theta \,,
\end{equation}
where
\begin{equation}\label{eq:entr2}
 \cW = I_4(\cH_0) - \left( 1 + I_4(G)\?\Jpar^2 \right)\?\Jpar^2\,.
\end{equation}
The physical scalars follow by constructing the symplectic section from \eqref{eq:ImO-attr} in the standard way as
\begin{equation}\label{eq:section-phys}
 2\? \ex^{-\u}\?\ex^{-\im\?\alpha} \cV =  -\frac1{2\?\sqrt{I_4(\sv_0)}}\? I^\prime_4(\sv_0) + \im \,\sv_0\,,
\end{equation}
and forming the ratios $t^i=X^i/X^0$.
Finally, the gauge field strengths are given by \eqref{eq:F-time-dec} and \eqref{eq:zeta-BPS} as
\begin{equation}\label{eq:F-attr-full}
 \mathsf{F} = d\left[ \left(- \frac12\? \ex^{4\?\u}\? I^\prime_4( \sv_0) - G \right)\,(r\?dt+\omega_0) \right] + \cF \,,
\end{equation}
where $\cF$ is given explicitly by \eqref{eq:F-attr}-\eqref{eq:attr-fin0}.

The physical quantities of interest include the conserved charges, electromagnetic and rotational, as well as the entropy associated to the black hole horizon.
The electromagnetic charges have been computed in \eqref{eq:attr-fin}, while for the computation of the entropy through the area law, it is useful to recast \eqref{eq:metr-bps-atrr} the following form, by completing the square with respect to $\phi$ and using \eqref{eq:e4u-atrr}
\begin{equation}\label{eq:metr-Sen}
ds^2_4 = \ex^{-2\?\u} \left( - r^2 \? d\tau^2 +   \frac{dr^2}{r^2} + \frac{ \vvv^2}{\Delta(\theta)}\? d\theta^2 \right)  
+ \ex^{2\?\u} \frac{\cW}{ \vvv^2} \?\Delta(\theta)\? \sin^2 \theta\? \?\left( d\phi + \frac{\Jpar} {\vvv\?\sqrt{\cW}}\?r\?d\tau\right)^2  \,.
\end{equation}
Here, we rescaled the time variable as $\tau = \vvv^2\? \cW^{-1/2}dt$ and $\cW$ is as in \eqref{eq:entr2}.
In the form \eqref{eq:metr-Sen}, it is easy to compute the black hole entropy $S$ through the horizon area $A$, via the Bekenstein-Hawking formula 
\begin{equation}\label{eq:BHentropy-horizon}
 S = \frac{A}{4} = \pi\?\sqrt{I_4(\cH_0) - \left( 1 + I_4(G)\?\Jpar^2 \right)\?\Jpar^2} \equiv \pi\? \sqrt{\cW}\,,
\end{equation}
where we have set $G_N = 1$ for simplicity.
The final conserved quantity is the angular momentum $\cJ$, which can be computed in terms of the appropriate Noether integral at the horizon. However, we will defer this discussion until section \ref{subsec:charges}, where both the Noether integral as well as the simpler Komar integral it reduces to in the asymptotic region are presented. Here, we simply cite the result (cf. \eqref{eq:J-general})
\begin{equation}\label{eq:J-general-0}
 \cJ = - \frac{\Jpar}{2}\?\left( \Iprod{I_4^\prime(G)}{I_4^\prime(\cH_0)} -\frac12\? I_4(\cH_0, \cH_0, G, G)\?\Iprod{G}{\cH_0} \rule[.1cm]{0pt}{\baselineskip}\right)\,,
\end{equation}
which should be solved together with \eqref{eq:attr-fin} for $\Jpar$ and $\cH_0$ in terms of $\cJ$ and $\Gamma$, in order to obtain the entropy in terms of conserved charges through \eqref{eq:BHentropy-horizon}, as discussed in section \ref{subsec:charges}.

\subsection{Examples}
\label{subsec:nhg-examples}
In this section, we present four examples of rotating BPS near-horizon geometries, each corresponding to one of the three interesting asymptotics listed in section \ref{sec:as-vac}. The attractor solution corresponding to an asymptotically AdS$_4$ flow within the T$^3$ model is new, and the full flow will be presented in the next section. The examples corresponding to flows with hvLif and Minkowski asymptotics are known in the literature, as the near-horizon regions of the KLR black hole \cite{Kunduri:2006ek} and of the non-BPS black hole of \cite{Bena:2009ev} respectively\footnote{Both of these solutions were constructed within 5d supergravity, here we consider their reduction to 4d along an angular isometry.}.

\subsubsection{Models with AdS\texorpdfstring{$_4$}{4} vacuum} \label{Model_ads_vac}

\subsubsection*{The T$^3$ model}
We now turn to the example of a rotating attractor in the simplest symmetric model in the cubic series, the T$^3$ model, with prepotential as in \eqref{eq:T3-model}. We focus on this model for brevity, noting that the extension to the STU model \eqref{eq:STU-def} and indeed to any cubic model is straightforward.

For the T$^3$ model, an AdS$_4$ vacuum exists in presence of mixed electric and magnetic gauging vector $G$, as
\begin{equation}\label{eq:ex1-G}
G = \{ g^0,\?0,\? 0,\? g_1 \}^{\rm T}\,,
\end{equation}
so that the components $g^1=g_0=0$ while $g^0$, $g_1$ are arbitrary.
For this gauging, we list the relevant objects
\begin{equation}\label{eq:ex1-I4G}
I_4(G)= -4\? g^0 g_1^3 \,, \qquad I^\prime_4(G) = \{ 0,-4\?g^0 g_1^2,4\? g_1^3,0\}^{\rm T}\,.
\end{equation}
This gauging leads to the cosmological constant $\Lambda = -3 \sqrt{I_4(G)}= -3 \sqrt{-4\? g^0 g_1^3}$, so that we choose to focus on the case $g^0<0$, $g_1 >0$ which gives $I_4(G)>0$ and a real and negative cosmological constant.

Similarly, we choose the charges carried by the black hole as 
\begin{equation}\label{eq:ex1-ch0}
\Gamma = \{ 0,\?p^1,\? q_0,\? 0 \}^{\rm T}\,,
\end{equation}
for simplicity. The constraint \eqref{eq:final-Sig} can be then implemented by trading one of the charges for $\kappa$, and we take
\begin{equation}\label{eq:ex1-ch}
\Gamma = \{ 0,\?p^1,\? \frac{\kappa+3\?g_1\? p^1}{g^0},\? 0 \}^{\rm T}\,.
\end{equation}

The complete attractor solution is based on the vector $\cH_0$ obtained as the solution to \eqref{eq:attr-fin}, for the gauging and charge vectors given above, and for an arbitrary constant $\Jpar$. The resulting vector also has only two nonzero components, as
\begin{equation}\label{eq:ex1-H}
\cH_0 = \{ 0,\?h^1,\? h_0,\? 0 \}^{\rm T}\,.
\end{equation}
The nontrivial components of \eqref{eq:attr-fin} are
\begin{equation}\label{eq:ex1-attr}
p^1 =  h^1\? (g_1 h^1 - h_0 g^0)- 2\? g^0 g^2\? \Jpar^2 \,, \qquad 
\frac{\kappa+3\?g_1\? p^1}{g^0} = h_0\?(h_0 g^0 + 3\? g_1 h^1) + 2\? g_1^3\? \Jpar^2\,,
\end{equation}
and their solution reads\footnote{Since the attractor equations always have the square of the symplectic vector $\cH_0$, its components are always determined upto an overall sign. Here and everywhere else in this paper we just present the solution with one chosen sign, since the other one is trivial to obtain and does not lead to a physically different solution. Additionally, there can be several branches of solutions due to different determination of square roots, and these do lead to potentially different solutions. In particular only the upper signs lead to regular spherical solutions in the solution discussed here. Depending on the range of charges, the cases with $\kappa = 0, -1$ do allow for both signs.}
\begin{align}\label{eq:ex1-attr-sol}
h^1 =&\, \frac{1}{4\?g_1}\?\left( \sqrt{\kappa + 12\?g_1\?p^1 + 16\?g_1^3 g^0 \?\Jpar^2} \mp \sqrt{\kappa + 4\?g_1\?p^1} \right)\,,
\\
\nonumber h_0 =&\, - \frac{1}{4\?g^0}\?\left( \sqrt{\kappa + 12\?g_1\?p^1 + 16\?g_1^3 g^0 \?\Jpar^2} \pm 3\? \sqrt{\kappa + 4\?g_1\?p^1} \right)\,.
\end{align}
Below, we will not use these explicit expressions in displaying results for simplicity, and continue to use the parameters $h^1$, $h_0$ instead of the charges, unless stated otherwise.

After setting up the above objects, we are now ready to display the various physical fields. The metric is of the form \eqref{eq:metr-bps-atrr}, with the metric along the sphere as in \eqref{eq:base-bps-sph}. Here, we focus on the physically more interesting case of a spherical horizon, so that we set $\kappa=1$ henceforth, noting that the cases with $\kappa=0,-1$ can be treated similarly. The function $\Delta(\theta)$ is given by \eqref{eq:Delta-sol} using \eqref{eq:ex1-I4G}, as
\begin{equation}\label{eq:ex1-delta_th}
\Delta(\theta) = 1 - 4\? g_1^3 g^0 \? \Jpar^2 \sin^2 \theta \,,
\end{equation}
while the constant $\vvv$ is fixed through \eqref{eq:vvv} as
\begin{equation}\label{eq:ex1-vvv}
\vvv =  3\? g_1 h^1 -h_0 g^0 \,.
\end{equation}
The scale factor of the metric is given by
\begin{equation}\label{eq:ex1-e4u-atrr}
\ex^{-4\?\u} = \frac1{\vvv^4}\?\left( 4\? (h^1)^3 h_0 - \Jpar^2 \?\left( 1 - 4\? g^0 g_1^3 \Jpar^2 \right) + \Jpar^2\? \Delta(\theta)\?  \sin^2 \theta \right)\,,
\end{equation}
consistent with \eqref{eq:e4u-atrr}.

The remaining physical fields may now be constructed using the formulae given in the previous section.
The scalar follows from solving \eqref{eq:ImO-attr}, with the result
\begin{align}
t = \frac12\?\frac{\im\? \vvv^2 \?e^{2\u} +\Jpar\? \cos \theta (h_0 g^0 +h^1  g_1 )}{ (h^1) ^2 -g_1 g^0\? \Jpar^2 \?\cos^2 \theta }\,,
\end{align}
while the gauge fields follow from \eqref{eq:F-attr-full} as
\begin{align}
A^0 = &\? -\frac{2\? (h^1) ^3- g^0\? \Jpar^2\? (h_0\? g^0 + 3\? h^1 g_1)\? \cos^2 \theta}{\vvv^{3}\?\ex^{-4\?\u}}\,(r\?dt+\omega_0) - g^0\? r\? dt
\nn\\[6pt]
&\? - \Jpar \?g^0 \?(h_0\? g^0 + 3\? h^1 g_1)\? \sin^2\theta \, d\phi  \,,
\end{align}
and 
\begin{align}
A^1 = &\? \Jpar\? \cos \theta \?\frac{h^1 \? (h_0 g^0 - h^1 g_1) + 2\?g^0 g_1^2 \, \Jpar^2\? \cos^2 \theta}{\vvv^{3}\?\ex^{-4\?\u}}\,(r\?dt+\omega_0)
\nn\\[6pt]
&\? - \left( p^1 + 2\?g^0 g_1^2 \, \Jpar^2\?\sin^2\theta \right)\?\cos \theta \, d\phi \,,
\end{align}
where in the last expression we used the charge $p^1$ in \eqref{eq:ex1-attr} for simplicity. We stress once again that in the spherical case one should use the upper sign in \eqref{eq:ex1-attr-sol} to obtain the expressions of all the fields in terms of the conserved charges only. We have checked explicitly that this solution satisfies the equations of motion for the $T^3$ model, as expected.

From the explicit form of the near horizon geometry we can read off the black hole entropy, which assumes a particularly simple expression in terms of the components of $\mathcal{H}$, as
\begin{equation}
S_{BH,\mathcal{H}} =\sqrt{ 4\? (h^1)^3 h_0 + \Jpar^2 \? (4\? g^0\? g_1^3\? \Jpar^2- 1)} \,,
\end{equation}
or alternatively, in terms of the electromagnetic charges:
\begin{equation}\label{eq:ex1-S}
S_{BH}= \sqrt{\frac{ (1+4\? g_1\? p^1)^{3/2} \sqrt{16\? g_1^3 g^0\? \Jpar^2+12\? g_1\? p^1+1} - \left( 24\? g_1^2\? (p^1)^2+12\? g_1\? p^1+1 \right) }{-8\? g_1^3 g^0}}\,.
\end{equation}
The latter expression reduces to the known entropy formula for the static black hole of \cite{Dall'Agata:2010gj, Hristov:2010ri} for $\Jpar=0 $, the parameter $\Jpar$ being related to the angular momentum. The precise relation of this parameter to the conserved angular momentum $\mathcal{J}$ will be given in section \ref{subsec:charges}, by means of the Komar integral in the the asymptotic region. We refer to that section for the expression for the entropy in terms of the conserved charges.

\subsubsection*{The $X^0 X^1$ model}
Let us also briefly discuss another model with an AdS$_4$ vacuum, the $X^0 X^1$ model where supersymmetric rotating hyperbolic black holes with an ergosphere were previously found \cite{Klemm:2011xw}.

For the $X^0 X^1$ model in the form \eqref{eq:X0X1-model}, an AdS$_4$ vacuum exists in presence of purely electric gauging vector $G$, as
\begin{equation}
G = \{0,\?0,\? g_0,\? g_1 \}^{\rm T}\,,
\end{equation}
so that the components $g^0=g^1=0$ while $g_0$, $g_1$ are assumed to be positive without loss of generality. The AdS$_4$ asymptotics are fixed by
\begin{equation}
 I_4(G)= (g_0 g_1)^2 \,, \qquad I^\prime_4(G) = 2 g_0 g_1\ \{ g_1, g_0, 0,0\}^{\rm T}\,,
\end{equation}
so that the asymptotic scalar is a positive constant fixed by the ratio of $g_0$ and $g_1$. 

We then choose a purely magnetic charge vector,
\begin{equation}
\Gamma = \{ p^0,\?p^1,\? 0,\? 0 \}^{\rm T}\,,
\end{equation}
and it then easily follows that the attractor solution given by the vector $\cH_0$ also needs to have purely magnetic components,
\begin{equation}\label{eq:ex4-H}
 \cH_0 = \{ h^0,\?h^1,\? 0,\? 0 \}^{\rm T}\,.
\end{equation}
The nontrivial components of \eqref{eq:attr-fin} then read
\begin{equation}
 p^0 =  g_1 \left(h^0 h^1+g_0 g_1 \Jpar^2 \right) \,, \qquad 
p^1 =  g_0 \left( h^0 h^1+g_0 g_1 \Jpar^2 \right)\,,
\end{equation}
which, together with the constraint for arbitrary internal space $\Iprod{G}{\Gamma} = -\kappa$, imply that
\begin{equation}
	g_0 p^0 = g_1 p^1 = - \frac{\kappa}{2}\ .
\end{equation}
One can easily see that in this special case the attractor mechanism cannot fully fix the scalars in terms of the charges, due to the fact that only the combination $h^0 h^1$ appears in the above equations. We can nevertheless show that a smooth horizon in this class must necessarily be of the hyperbolic type, by evaluating the entropy from \eqref{eq:BHentropy-horizon},
\begin{equation}
	S = - \frac{\pi\ \kappa}{2 g_0 g_1}\ ,
\end{equation}
which is only positive in the hyperbolic case $\kappa = -1$.

We have therefore constructed a rotating hyperbolic black hole near-horizon geometry that is a smooth continuation of the static case. This appears to be in another BPS branch of rotating solutions compared to the one discovered in \cite{Klemm:2011xw} in the same $X^0 X^1$ model.

\subsubsection{Models from 5d reduction}

We now turn to an example attractor in a model that does not admit an AdS$_4$ vacuum, but exhibits runaway behaviour. Nevertheless, this model is physically relevant, since it arises from dimensional reduction of the gauged STU model in five dimensions, which admits an AdS$_5$ vacuum. Indeed, reduction of an asymptotically AdS$_5$ spacetime along an isometry leads to a metric with hvLif asymptotics of the type \eqref{eq:rank3-vec}. It follows that one can - in principle - construct physically interesting solutions in five dimensions by lifting solutions of models with gaugings vectors $G$ of rank-3, as in \eqref{eq:metr-rank-3}. Here, we do not pursue such a goal, restricting ourselves to matching an attractor constructed in a model with rank-3 gauging to the reduction of the known solutions of \cite{Gutowski:2004ez,Gutowski:2004yv, Kunduri:2006ek}.

Our starting point is the rank-3 vector of FI terms
\begin{equation}
G = \sqrt{2}\?\{ 0,\?0,\?0,\?0,\? g_0,\? g,\? g,\? g \}^{\rm T} \,,
\end{equation}
and we consider a set of charges of the type
\begin{equation}\label{eq:ex2-ch}
\Gamma = \{ p^0,\?0,\? q_0,\? q_i \}^{\rm T}\,.
\end{equation}
In order to connect to the four parameters of the solution in \cite{Kunduri:2006ek}, which can be thought of as three electric charges and one angular momentum, we fix the five charges in \eqref{eq:ex2-ch} in terms of four parameters, $\delta$ and $\mu_1$, $\mu_2$, $\mu_3$ as in \eqref{eq:KLR-charge}, so that there is one constraint among them, whose explicit form is not important for the following. The condition \eqref{eq:final-Sig} now leads to the identification
\begin{equation}
 g_0 = \cosh\delta\,,
\end{equation}
and one can verify that the attractor equation \eqref{eq:attr-fin} is solved by the vector
\begin{align}\label{eq:ex2-sol}
 \cH_0 =&\,\frac{\sqrt{2}\? \vvv}{8\?g^3}\? \{ 0,\? 2\?g ,\? 2\?g ,\? 2\?g ,\?\tfrac12 - g_0,\? - g_0\?g,\? - g_0\?g,\? - g_0\?g \}^{\rm T} 
\nn\\
&\, + \frac1{\sqrt{2}\?g_0\? \vvv}\? \{ 1,\? - m_1 ,\? - m_2 ,\? - m_3 ,\? m_1\?m_2\?m_3,\?  m_2\?m_3 ,\?m_1\?m_3,\? m_1\?m_2 \}^{\rm T}\,,
\end{align}
where we defined the shorthand parameters
\begin{equation}
 m_i = \frac{1+g^2\?\mu_i}{2\?g}\,, 
\end{equation}
and the remaining parameters of the solution are given as
\begin{equation}\label{eq:ex2-J}
\vvv^2 = 2\?\frac{g}{g_0}\?(m_1 + m_2 + m_3) - 2 \,. \qquad \Jpar = -\frac{\sinh{\delta}}{8\?g^3}\? \vvv \,.
\end{equation}
Note that \eqref{eq:ex2-J} fixes $\Jpar$ in terms of the other parameters, since the four free parameters of \cite{Kunduri:2006ek} have been interpreted as charges in the 4d solution above. The situation is reversed in \cite{Kunduri:2006ek}, as there are three electric charges in the 5d solution and the fourth parameter corresponds to the angular momentum parameter $\Jpar$ above, while $p^0$ is interpreted as a geometrical parameter in five dimensions and is therefore fixed in terms of the other quantities.

We refer to Appendix \ref{sec:klr} for a review of the solution of \cite{Kunduri:2006ek}, its recasting in terms of the parameters above and the reduction to four dimensions. Comparison with \eqref{eq:KLR-J} and \eqref{eq:KLR-G} shows that the attractor described by \eqref{eq:ex2-sol}-\eqref{eq:ex2-J} indeed matches with the result of this reduction near the horizon. The expression in \eqref{eq:KLR-sol} allows to construct the complete four dimensional solution, which can be seen to asymptote exactly to the hvLif geometry \eqref{eq:metr-rank-3} for $\delta=0$, while for $\delta\neq 0$ one finds a more general rotating asymptotic geometry with the same hvLif exponents.

\subsubsection{Asymptotically flat black hole}

Finally, we present an example of an attractor corresponding to an asymptotically flat black hole. Since the asymptotic Minkowski vacuum is not supersymmetric in an abelian gauged theory, such black holes can be viewed both as flows between this vacuum and a supersymmetric AdS$_2$ \cite{Hristov:2012nu} and as non-BPS black holes in ungauged supergravity \cite{Bossard:2012xsa}.\footnote{Since the potential vanishes identically in this case, the bosonic sector reduces to that of ungauged supergravity.}

A concrete example is given by the STU model with FI parameters as in \eqref{eq:rank1-vec}, explicitly
\be
g^0 = g^i = g_i = 0 \qquad g_0\neq 0 \,,
\ee
which leads to a Minkowski vacuum. Choosing the charges as
\begin{equation}\label{eq:ex3-ch}
\Gamma = \{ p^0,\?0,\? 0,\? q_i \}^{\rm T}\,,
\end{equation}
this model admits an asymptotically flat under-rotating black hole solution whose near horizon geometry solves the BPS equation we presented. The constraint \eqref{eq:final-Sig} together with the attractor equation \eqref{eq:attr-fin} can be solved as
\begin{equation}
 g_0 = -\frac{1}{p^0}\,, \qquad \vvv=1\,, \qquad \cH_0 = - \{ p^0,\?0,\? 0,\? \frac{\sqrt{p^0 q_1 q_2 q_3}}{q_i} \}^{\rm T}\,,
\end{equation}
while \eqref{eq:Delta-sol} collapses to $\Delta(\theta)=1$. 
The metric then assumes the standard form for under-rotating attractors in ungauged supergravity:
\be
ds^2 = -e^{2\?\u} \left(r\?dt+\omega_0 \right)^2 + e^{-2\?\u} \left( \frac{dr^2}{r^2} + d\theta^2 + \sin^2 \theta d\phi^2 \right)
\ee
with 
\be
e^{-4\?\u} = - 4 p^0 q_1 q_2 q_3 - \Jpar^2 \cos^2 \theta \qquad \quad
\omega_0 =  \Jpar \sin^2 \theta\? d\phi\,.
\ee 
The remaining fields take the form summarized in section \ref{sec:summ-attr} and are given explicitly in \cite[App. C]{Hristov:2012nu}.

\section{Full rotating flow in AdS\texorpdfstring{$_4$}{4}}
\label{sec:full-sol}

\subsection{The rotating black hole solution}
\label{sec:flow}
Here we consider the full BPS flow for rotating black holes, interpolating between the attractor solution in the previous section and asymptotically locally AdS$_4$. Inspired by the known solutions in the static case \cite{Cacciatori:2009iz,Dall'Agata:2010gj}, in the form given in \cite{Katmadas:2014faa, Halmagyi:2014qza} and reviewed in section \ref{sec:static-bh}, we take the three-dimensional base metric as in \eqref{eq:3d-metr} and concentrate for simplicity on the case where $\Sigma$ is of spherical topology, as in \eqref{eq:base-bps-sph}, so that the base metric becomes
\begin{equation}
 ds^2_3 = dr^2 + \ex^{2\?\psi(r)} \left( \frac{d\theta^2}{\Delta(\theta)} + \?\Delta(\theta)\?\sin^2\!\theta\? d\phi^2 \right)\,.
\end{equation}
The total metric \eqref{eq:metr-bps} then takes the explicit form\footnote{In section 5 of \cite{Gnecchi:2013mja} an ansatz for rotating AdS$_4$ black holes with arbitrary prepotentials was put forward. The analysis presented here suggests a possible generalization of the form of the function $v$, which in (5.2) of \cite{Gnecchi:2013mja} reads $v = Q-P$, to $v = \xi_1 Q - \xi_2 P$, with $\xi_1,\xi_2$ constant. In particular, for the specific case treated in this section we choose $\xi_2=0$, and this enables us to find the novel analytic solutions for the full flow for arbitrary models with vector multiplets.}
\begin{equation}\label{eq:metr-bps-full}
ds^2_4 = -\ex^{2U} \?(dt + \omega )^2 + \ex^{-2U}\?\left(  dr^2 + \frac{ \ex^{2\?\psi} }{\Delta(\theta)}\? d\theta^2 +  \ex^{2\?\psi}\?\Delta(\theta)\?\sin^2\!\theta\? d\phi^2 \right)  \,.
\end{equation}
We now proceed to solve the BPS conditions \eqref{eq:final} for the full flow, following similar steps as in section \ref{subsec:nhg-BPS}. To this end, we introduce the combination 
\begin{equation}\label{eq:ItoH}
 \ex^\psi \? \sv = \cH (r) +  \Jpar \cos \theta\ G\ ,
\end{equation}
which allows to write the flow equation \eqref{eq:final-flow} as
\begin{align}\label{eq:fl-Ortin-fin}
\ex^{-\psi}\?\star d (\cH +  \Jpar \cos \theta\ G)  -\frac1{4}\? \ex^{-2\?\psi}\?I^\prime_4( \cH +  \Jpar \cos \theta\ G , \cH  +  \Jpar \cos \theta\ G, \star\hat{G}) - r\?d\omega\? G + \cF = 0\,,
\end{align}
while the remaining BPS equations in \eqref{eq:final} become 
\begin{align}
   \Iprod{ G }{\cH}  =&\, ( \ex^\psi )^\prime \,, \label{eq:base-fin}
  \\   
  \Iprod{\cal A}{G} = &\, \frac{\left( \Delta(\theta)\sin^2{\theta}  \right)^\prime}{2\?\sin{\theta}}\? d\phi\? \,, \label{eq:gauge-fin}
\\
 \ex^{2\?\psi} \star d\omega = &\, \Iprod{  d (\cH +  \Jpar \cos \theta\ G ) }{ \cH +  \Jpar \cos \theta\ G } \nonumber \\
& +\frac{1}{6}\?\?\ex^{-\?\psi} I_4( \cH +  \Jpar \cos \theta\ G , \cH+  \Jpar \cos \theta\ G , \cH+  \Jpar \cos \theta\ G , G )\?dr\,. \label{eq:domega-fin}
\end{align}
This set of equations is a direct generalization of the corresponding static one, so that we may be guided by the known solutions to it. 

Here, we restrict for brevity on generalizing the simpler class of \cite{Katmadas:2014faa} to the rotating case, expecting that the most general solution of the static equations in \cite{Halmagyi:2014qza} can be also treated along the same lines. We therefore adopt the simple ansatz
\begin{equation}\label{eq:flow-ansatz}
\cH (r) = \cH_0 + \cH_\infty\ r  \,,   \qquad   \omega =  \left( \ominf(\theta)  -\Jpar\? \frac{\Delta(\theta)}{\ex^{\psi}} \right) \?\sin^2\!\theta\?d\phi \,,
\end{equation}
where the two constant symplectic vectors $ \cH_0, \cH_\infty$ and the function $\ominf(\theta)$ are to be determined. In this form, it is manifest that this ansatz reduces to the attractor solution of the previous section for $r\rightarrow 0$, while $\cH_\infty$ and $\ominf$ parametrize the asymptotic region. Inserting \eqref{eq:flow-ansatz} in \eqref{eq:base-fin} leads to the following expression for the function $\ex^\psi$, as
\begin{align}\label{eq:epsi-full}
 \ex^\psi = &\, \frac12\? \Iprod{G}{\cH_\infty}\?r^2 + \Iprod{G}{\cH_0}\?r  =  I_4(G)^{1/4}\?r^2 + \Iprod{G}{\cH_0}\?r  \,,
\end{align}
where we disregarded an additive integration constant and in the second equality we imposed that the $\cO(r^2)$ term is such that \eqref{eq:metr-bps-full} allows for AdS$_4$ asymptotics. Similarly, the BPS flow equation \eqref{eq:fl-Ortin-fin} reduces to
\begin{align}\label{eq:fl-Ortin-ans}
&\?\ex^{\psi}\?\cH_\infty\?\sin\theta\?d\theta\wedge d\phi  
 -\frac1{4}\? I^\prime_4( \cH_0 + \Jpar\? \cos\theta\?G , \cH_0 + \Jpar\? \cos\theta\?G , G)\?\sin\theta\?d\theta\wedge d\phi 
\nonumber\\
&\? \qquad\qquad\qquad\qquad
- r\?d( \ominf\?\sin^2\!\theta)\wedge d\phi\, G + d\left(\Jpar\? r\?\frac{\Delta(\theta)}{\ex^\psi} \?\sin^2\!\theta\?d\phi \right)\? G + \cF = 0\,,
\end{align}
with $\ex^{\psi}$ as in \eqref{eq:epsi-full}.

Note that since $\ominf$ depends only on $\theta$, only the last two terms in \eqref{eq:fl-Ortin-ans} can have a leg along $dr$. It then follows that we can solve this flow equation order by order in $r$ for the $d\theta \wedge d \phi$ terms, similar to the static case.
Starting with the ${\cal O}(r^2)$ terms of \eqref{eq:epsi-full}-\eqref{eq:fl-Ortin-ans} we find that they are solved by fixing the vector $\cH_\infty$ as
\begin{align}\label{eq:H-infty-sol}
\cH_\infty = \frac{1}{2}\? I_4(G)^{-3/4}\?I_4(G)^{\prime}\,,
 \end{align}
in exactly the same way as in the static case.
The sub-leading term in $r$ instead reduces to
\begin{align}
 \Iprod{G}{\cH_0}\? \cH_\infty = \frac1{2}\? I^\prime_4( \cH_\infty ,\cH_0 + \Jpar\? \cos\theta\?G, G) + \frac{\partial_\theta( \ominf\?\sin^2\!\theta)}{\sin\theta}\? G \,,
\end{align}
which can be further simplified using \eqref{eq:proj}, leading to the conditions
\begin{gather}
 \Iprod{\cH_0}{\cH_\infty}=0\,\quad \Rightarrow \quad I_4 (\cH_0, G, G, G) = 0\ , \label{eq:flow-conds}\\
 \ominf = \Jpar\? I_4(G)^{1/4}\,.
\end{gather}
With these results, the flow equation \eqref{eq:fl-Ortin-ans} reduces to the following expression for the gauge field strengths
\begin{align}\label{eq:gauge-fields-sol}
\cF = \frac1{4}\? I^\prime_4( \cH_0 + \Jpar\? \cos\theta\?G , \cH_0 + \Jpar\? \cos\theta\?G , G)\?\sin\theta\?d\theta\wedge d\phi - d\left(\Jpar\? r\?\frac{\Delta(\theta)}{\ex^\psi} \?\sin^2\!\theta\?d\phi \right)\? G \,,
\end{align}
which is manifestly closed and generalizes the attractor fluxes \eqref{eq:attr-fin0}-\eqref{eq:F-attr} to the full flow. With these results, it is straightforward to verify that \eqref{eq:gauge-fin} and \eqref{eq:domega-fin} reduce to their attractor counterparts, leading to 
 \begin{empheq}[box=\fbox]{align}\label{eq:final-sph-2}
	\Iprod{G}{\Gamma} = -1\ ,
\end{empheq}
due to our choice of spherical topology, as well as \eqref{eq:Delta-sol} and \eqref{eq:NUT-ch}, which determine the function $\Delta(\theta)$ in \eqref{eq:flow-ansatz} and ensure regularity, respectively.

Integrating the field strengths $\cF$ in \eqref{eq:gauge-fields-sol} along the sphere leads to the attractor equation
 \begin{empheq}[box=\fbox]{align}\label{eq:attr-fin-2}
  \Gamma = \frac{1}{4}\? I^\prime_4\left(\cH_0, \cH_0, G \right) + \frac{1}{2}\? \Jpar^2\? I^\prime_4\left( G \right)\,,
\end{empheq}
which, as expected, is identical to the attractor condition in \eqref{eq:attr-fin}, upon identifying the vector $\cH_0$ with the attractor solution. Finally we also find the constraints
 \begin{empheq}[box=\fbox]{align}\label{eq:charge-constraints}
	I_4 (\Gamma, \Gamma, \Gamma, G)  = I_4 (G, G, G, \Gamma)  = 0\ ,
\end{empheq}
upon contracting \eqref{eq:attr-fin-2} with $I_4'(G)$ and using \eqref{eq:flow-conds}. One can write the solution to \eqref{eq:attr-fin-2}-\eqref{eq:charge-constraints} by using the explicit solution of \cite[Sec. 3.2]{Halmagyi:2014qza} in the static case, for a shifted charge $\hat\Gamma = \Gamma - \frac{1}{2}\? \Jpar^2\? I^\prime_4( G )$, as
\begin{equation}\label{eq:H0-sol}
 \cH_0 = \frac{1}{\sqrt{C_1}}\?\left( C_2\? I^\prime_4( G ) + I^\prime_4(\hat\Gamma, \hat\Gamma, G) \right)\,,
\end{equation}
where the two constants $C_1$, $C_2$ are given by
\begin{align}
C_1(\hat\Gamma,G) = &\? C_1(\Gamma,G) =\Iprod{G}{\Gamma}\?I_4(\Gamma, \Gamma, G, G)-2\? \Iprod{I_4^\prime(G)}{I_4^\prime(\Gamma)} \,,
\\
C_2(\hat\Gamma,G) = &\? -\frac{1}{2\?I_4(G)}\,\left( \frac14\?I_4(\hat\Gamma, \hat\Gamma, G, G) \pm \sqrt{\frac1{16}\?I_4(\hat\Gamma, \hat\Gamma, G, G)^2 - 4\?I_4(G)\? I_4(\hat\Gamma) } \right) \,,
\label{eq:H0-sol-coeff}
\end{align}
and we stress that the shift in $\Gamma$ cancels out in the expression for $C_1$, a property that will be important below.

This concludes the construction of the BPS black hole flows connecting the family of attractors in the previous section to asymptotic AdS$_4$. Summarizing, one starts from a charge vector satisfying the constraints \eqref{eq:final-sph-2} and \eqref{eq:charge-constraints} for a given given gauging $G$, and solves \eqref{eq:attr-fin-2} for the vector $\cH_0$. The metric for the solution is then given by \eqref{eq:metr-bps-full} with the scale factor $\ex^\psi$ in  \eqref{eq:epsi-full} and the one-form $\omega$ in \eqref{eq:flow-ansatz}, while the function $\Delta(\theta)$ is given by \eqref{eq:Delta-sol}. The second scale factor $\ex^U$ is given by the expression 
\begin{align}\label{eq:eU-flow}
 \ex^{-4U} = &\? \ex^{-4\psi}\left(I_4(\cH) +\frac14\? I_4(\cH, \cH, G, G)\? \Jpar^2 + I_4(G)\? \Jpar^4   \rule{0pt}{6mm} \right.
\nn\\
&\? \left.\rule{0pt}{6mm} \hspace{12mm} + \left( \Delta(\theta) -2  \?  I_4(G)^{1/4}\ex^\psi \right) \Jpar^2 \sin^2\theta  \right) \,,
\end{align}
obtained from \eqref{eq:ItoH}, with $\cH$ as in \eqref{eq:flow-ansatz}. The scalar fields are also given by \eqref{eq:ItoH}, as explained around \eqref{eq:section-phys}, upon replacing $\sv_0$ by $\sv$ in \eqref{eq:ItoH}.
Finally, the gauge field strengths are given by
\begin{equation}\label{eq:F-full}
 \mathsf{F} = d\left[ \left(- \frac12\? \ex^{4\?U}\? I^\prime_4( \sv) - G\?r \right)\,(dt+\omega) \right] + \cF \,,
\end{equation}
where the spatial components $\cF$ are given in \eqref{eq:gauge-fields-sol}.

\subsection{Conserved charges}
\label{subsec:charges}

We now turn to a discussion on the physical properties of the full black hole solutions constructed in section \ref{sec:flow}. As mentioned above, the electromagnetic charge vector is computed through its standard definition \eqref{eq:charge-def} with the result \eqref{eq:attr-fin-2}, which also constitutes the attractor equation for the scalars. 

Turning to the conserved charges in the gravitational sector, it is useful to note that the superalgebra underlying the BPS black hole solutions we have described is $U(1|1)$ \cite{Hristov:2011ye,Hristov:2011qr}. The latter is characterized by the anticommutation relation between two supercharges $ \{Q^I,Q^J \} = H \delta^{IJ}$, where $H$ is the generator of time translations, so that the angular momentum does not enter into the BPS bound. We therefore conclude that the mass of these solutions is $M=0$ since they saturate the BPS bound, while the angular momentum charge is free and has to be computed independently.

The computation of the conserved angular momentum proceeds through the Noether integral associated to diffeomorphisms along the angular isometry. The central object in this approach is the Noether potential, which for a two-derivative Lagrangian ${\cal L}$ containing the Einstein--Hilbert term as well as Maxwell and scalar fields minimally coupled to gravity, reads
\begin{equation}\label{eq:Noether-pot}
 Q_{\mu\nu}(\xi) = \nabla_\mu \xi_\nu + (\xi\cdot A^I)\?\frac{\partial {\cal L}}{\partial F^I{}^{\mu\nu}} 
 = \nabla_\mu \xi_\nu + (\xi\cdot A^I)\?G_I{}_{\mu\nu}\,,
\end{equation}
where $\xi_\mu$ is a Killing vector and in the second equality we used the definition of the dual gauge field strengths. The supergravity action \eqref{Ssugra4D} clearly falls in this class, so that we can construct the corresponding Noether integral for the angular momentum\footnote{Note that one can in principle also compute the mass using the Noether potential for the timelike isometry, but for asymptotically AdS spacetimes this integral diverges naively, and an appropriate renormalization is required.} based on \eqref{eq:Noether-pot} for the rotational Killing vector $\xi^\phi=\partial_\phi$. 

In particular, one can exploit the compactness of the orbits of the rotational Killing vector $\xi^\phi$ to write the angular momentum $\cJ$ as
\begin{equation}\label{eq:J-Noether}
 \cJ = \frac1{16\?\pi}\?\int_S dS^{\mu\nu} Q_{\mu\nu}(\xi^\phi) 
 = \frac1{16\?\pi}\?\int_S dS^{\mu\nu} \left( \nabla_\mu \xi^\phi_\nu + (A^I_\phi)\?G_I{}_{\mu\nu} \right) \,,
\end{equation}
where $S$ is any surface enclosing the black hole horizon and $dS^{\mu\nu}$ is its surface element. The conservation of the Noether potential \eqref{eq:Noether-pot} ensures that \eqref{eq:J-Noether} is independent of the surface $S$, so that one may compute this integral either at the attractor described in section \ref{subsec:nhg-BPS} or in the asymptotic region of the full black hole solutions in section \ref{sec:flow}. We have performed both computations, obtaining the same result as expected, but in this section we present in some detail only the computation in the asymptotic region, for brevity.

When $S$ in \eqref{eq:J-Noether} is identified with the asymptotic $\text{S}^2_{\infty}$, the asymptotic constants of the gauge fields $A^I_\phi$ can be set to zero by a judicious gauge transformation\footnote{This seems as an inherently asymptotic operation, but it can be enforced at the horizon by choosing the vector potentials $\cA$ corresponding to the horizon field strengths \eqref{eq:F-attr} such that $\cA(\theta=0)=-\cA(\theta=\pi)$, as in the standard prescription of \cite[Sec. 3]{Astefanesei:2006dd}.}, so that the integral collapses to the standard Komar integral
\begin{equation}\label{eq:J-Komar}
 \cJ = \frac1{16\?\pi}\?\int_{\text{S}^2_{\infty}} dS^{\mu\nu} \nabla_\mu \xi^\phi_\nu \,.
\end{equation}
In order to compute this integral explicitly, it is useful to put the metric in the canonical ADM form, which reads:
\begin{equation}\label{eq:metric-ADM}
ds^2 = - N^2 dt^2 + \sigma (d \phi - \Omega dt)^2 + \frac{dr^2}{Q} + \frac{d\theta^2}{T}\,,
\end{equation}
where the warp functions $N$, $\sigma$ and $\Omega$ are identified as
\begin{align}
 Q =&\? \ex^{2U}\,, \qquad  T= \ex^{2U} \ex^{-2\?\psi} \Delta(\theta) \,,
 \nonumber\\
 \sigma = &\? \ex^{-2\?U} \ex^{2\?\psi} \Delta(\theta)\? \sin^2 \theta - \ex^{2\?U} \omega^2\,,
 \label{eq:ADM}\\
 \Omega = &\? \frac{\ex^{2\?U} \omega}{\sigma} = \frac{\ex^{2\?U} \omega }{\ex^{-2\?U} \ex^{2\?\psi} \Delta(\theta)\? \sin^2 \theta - \ex^{2\?U} \omega^2}\,,
 \nonumber\\
 N^2 = &\? \ex^{2\?U} + \sigma\? \Omega^2  = \frac{\ex^{2\?\psi} \Delta(\theta) \? \sin^2 \theta}{\ex^{-2\?U} \ex^{2\?\psi} \Delta(\theta) \? \sin^2 \theta - \ex^{2\?U} \omega^2}\,. \nonumber
\end{align}
From this we can read off the angular velocity at infinity:
\begin{equation} \label{omega_inf}
\Omega_{\infty} = \lim_{r \rightarrow \infty} \Omega = I_4(G)^{3/4} \, \Jpar   = \frac{\Jpar}{l_{AdS}^3}\,.
\end{equation}
where $l_{AdS}^2 = I_4(G)^{-1/2}$. We note in passing that the angular velocity at the horizon, computed as $\Omega_h =\Omega(r=r_{h}) $ vanishes:
\begin{equation}
\Omega_h =0\,,
\end{equation}
as one may also verify by \eqref{eq:metr-Sen}.

In terms of the objects in \eqref{eq:ADM}, the angular momentum Komar integral \eqref{eq:J-Komar} is computed by using 
\begin{equation}
  \xi^\phi= \frac{\partial}{\partial\phi} \,, \qquad 
dS^{\mu\nu} = (v^{\mu} u^{\nu} - v^{\nu}u^{\mu})\,\sqrt{ \frac{\sigma}{T} }\?d\theta d\phi\,,
\end{equation}
where the ratio $\sigma/T$ denotes the induced metric on a two-sphere of constant $r$ and $t$, and the two vectors $u^{\mu}$, $v^{\mu}$ are given by
\begin{equation}
  u=\frac1N\?\left(\frac{\partial}{\partial t}+\Omega\?\frac{\partial}{\partial\phi} \right) \,, \qquad  v= \sqrt{Q}\,\frac{\partial}{\partial r}\,,
\end{equation}
i.e. they the normal vector of a $t$-constant hypersurface and the normal outward-pointing vector to the boundary, respectively.
Evaluation of \eqref{eq:J-Komar} proceeds straightforwardly, upon using the expressions in \eqref{eq:epsi-full}, \eqref{eq:flow-ansatz} and \eqref{eq:eU-flow} for $\ex^\psi$, $\omega$ and $\ex^U$ respectively in \eqref{eq:ADM}. The result reads
\begin{equation}\label{eq:J-general}
 \cJ = - \frac{\Jpar}{2}\?\left( \Iprod{I_4^\prime(G)}{I_4^\prime(\cH_0)} - \frac12\?I_4(\cH_0, \cH_0, G, G)\?\Iprod{G}{\cH_0} \rule[.1cm]{0pt}{\baselineskip}\right)\,,
\end{equation}
where we also used the conditions \eqref{eq:flow-conds} in the derivation. 

The expression \eqref{eq:J-general} for the angular momentum must be solved along with \eqref{eq:attr-fin-2} for $\cH_0$ and $\Jpar$ in order to obtain all physical quantities in terms of the conserved charges. Using the explicit solution \eqref{eq:H0-sol} for the $\cH_0$ along with the constraints \eqref{eq:charge-constraints}, one can show that \eqref{eq:J-general} takes the form
\begin{equation}\label{eq:J-Q}
 \cJ = \frac{\Jpar}{2}\?\sqrt{\Iprod{G}{\Gamma}\?I_4(\Gamma, \Gamma, G, G)-2\? \Iprod{I_4^\prime(G)}{I_4^\prime(\Gamma)} }   \,,
\end{equation}
which  is independent of $\Jpar$, as mentioned below \eqref{eq:H0-sol-coeff}.
It follows that one may use \eqref{eq:J-Q} to write $\Jpar$ in \eqref{eq:attr-fin-2} in terms of the charges, as 
 \begin{empheq}[box=\fbox]{align}\label{eq:attr-fin-J}
  \frac{1}{4}\? I^\prime_4\left(\cH_0, \cH_0, G \right) = \Gamma - \frac{2\?\cJ^2}{\Iprod{G}{\Gamma}\?I_4(\Gamma, \Gamma, G, G)-2\? \Iprod{I_4^\prime(G)}{I_4^\prime(\Gamma)}}\? I^\prime_4\left( G \right)\,,
\end{empheq}
and then obtain the solution to this equation for the $\cH_0$ by replacing the shifted charge $\hat\Gamma$ in \eqref{eq:H0-sol} for the shifted charge in the right hand side of \eqref{eq:attr-fin-J}. The entropy formula \eqref{eq:BHentropy-horizon} then takes the following form in terms of conserved charges only
\begin{equation} \label{entropy_bigJ-gen}
 S_{BH} 
= \pi  \sqrt{\frac{l_{AdS}^{4} }{2}\,\left( \frac14\?I_4(\Gamma, \Gamma, G, G) \pm \sqrt{\frac1{16}\?I_4(\Gamma, \Gamma, G, G)^2 - 4 \frac{( I_4(\Gamma) + \cJ^2)}{l_{AdS}^{4}} } \right)}\,,
\end{equation}
where $l_{AdS}^2 = I_4(G)^{-1/2}$ and we have again set $G_N = 1$. In the spherical case one needs to strictly choose the positive sign in the above formula in order to possibly find a positive real answer, i.e.\ a regular black hole. One can see that the angular momentum is bounded from above due to the term in the inner square root and that \eqref{entropy_bigJ-gen} reduces to the entropy formula of \cite{Halmagyi:2013qoa} for static black holes upon setting $\mathcal{J} =0$. 

Before concluding this section, we would like to make a few comments on the conformal boundary of our solutions, which pertain to the class of asymptotically locally AdS$_4$ spacetimes (see for example \cite{Papadimitriou:2005ii}). Taking the limit $r\rightarrow \infty$ of the metric \eqref{eq:metric-ADM}, we see that the boundary metric approaches the form
\begin{equation} \label{as_bound}
ds^2= r^2 \Delta(\theta) \left[- \frac{dt^2}{l_{AdS}^2} +\frac{d \theta^2}{\Delta(\theta)^2}+ \frac{\sin^2 \theta}{\Delta(\theta)} \left( d \phi +  \frac{\Jpar}{l_{AdS}^3} dt\right)^2  \right]\,.
\end{equation}
This is not the
standard metric on $R \times S^2$, due to the fact that there is a non-zero angular velocity at infinity $\Omega_{\infty}$, as in \eqref{omega_inf}. The metric in square bracket is that of the Einstein space $R \times S^2$ seen by a rotating frame of reference. The coordinate change
\begin{equation}
t' = \frac{t \sqrt{\Xi}}{l_{AdS}}\,, \qquad  \phi' = \phi \sqrt{\Xi} +  \frac{\Jpar}{l_{AdS}^2}  t' \,,  \qquad \Xi = 1 + \frac{\Jpar^2}{l_{AdS}^4}\,,
\end{equation}
brings the metric to the form
\begin{equation} \label{as_bound2}
ds^2= \frac{r^2 \Delta(\theta)}{\Xi} \left[-dt'^2+\frac{ \Xi d \theta^2}{\Delta(\theta)^2}+ \frac{\sin^2 \theta}{\Delta(\theta)}d \phi'^2  \right]\,,
\end{equation}
while the further reparametrization $\Xi \tan^2\theta' = \tan^2 \theta $ yields
\begin{equation}
ds^2 = r^2 \frac{\cos^2\theta}{\cos^2 \theta'} \left[- dt'^2 + d\theta'^2 + \sin^2 \theta' d\phi^2 \right]\,,
\end{equation}
which is the standard form $R \times S^2$ up to the conformal factor $ \cos^2 \theta/ \cos^2 \theta'$. Hence the boundary metric in \eqref{as_bound} is conformal to the standard boundary of four-dimensional AdS space. More details can be found for instance in  \cite{Hawking:1998kw,Papadimitriou:2005ii,Gibbons:2004ai,Gibbons:2005vp,Toldo:2016nia}. 
Note that the boundary data falls into the general class discussed in \cite{Hristov:2013spa}. After a Wick rotation to Euclidean signature, this case was studied in \cite{Benini:2015noa} and \cite{Closset:2018ghr} corresponding to the refinement of the topologically twisted index by angular momentum. 

\subsection{An example solution}
\label{sec:example-flow}

We now turn to an example rotating black hole solution in AdS$_4$ in the T$^3$ model, which is the full flow corresponding to the attractor in section \ref{Model_ads_vac}. The solution is naturally described by replacing the constants $\cH_0$ parameterizing the attractor in \eqref{eq:ex1-H} by the linear functions $\cH$ in \eqref{eq:flow-ansatz}, as 
\begin{gather}\label{eq:exF-H}
\cH = \{ 0,\?\cH^1,\? \cH_0,\? 0 \}^{\rm T}\,,
\nn\\
\cH^1 = h^1 + \frac1{\sqrt{2}}\?\left(-\frac{g^0}{g_1} \right)^{1/4} r\,, \qquad \cH_0 = h_0 + \frac1{\sqrt{2}}\?\left(-\frac{g_1}{g^0} \right)^{3/4} r\,.
\end{gather}
The scale factor $\ex^\psi$ resulting from \eqref{eq:epsi-full} reads
\begin{equation}
\ex^\psi = (-4\? g^0 g_1^3)^{1/4}r^2 + (- h_0 g^0 + 3\? g_1 h^1)\?r \,,
\end{equation}
while the scale factor $\ex^U$ is given concisely as 
\begin{align}
\ex^{-4U} = &\? \ex^{-4\psi}\left( 4\?\cH_0\?(\cH^1)^3 - \left( (\cH_0 g^0)^2 + 6\?  g^0 g_1 \cH_0 \cH^1 -3 (g_1\? \cH^1)^2 \right)\? \Jpar^2 -4\? g^0 g_1^3\? \Jpar^4  \rule{0pt}{6mm} \right.
\nn\\
&\? \left.\rule{0pt}{6mm} \hspace{12mm} + \left( \Delta(\theta) -2  \? (-4\? g^0 g_1^3)^{1/4}\ex^\psi \right) \Jpar^2 \sin^2\theta  \right) \,,
\end{align}
where $\Delta(\theta)$ is as in \eqref{eq:ex1-delta_th}. The rotation one-form $\omega$ is given by
\begin{equation}
\omega =  \Jpar\? \left(\rule{0pt}{4mm} (-4\? g^0 g_1^3)^{1/4}  - \ex^{-\psi}\ \Delta(\theta) \right) \?\sin^2\!\theta\?d\phi \,.
\end{equation}

With these definitions, the complex scalar assumes the form
\begin{equation}
z= \frac12\? \frac{ \Jpar\? \cos \theta\? ( g_1\? h^1 +g^0\? h_0) + \im\? \sqrt{4\?\cH_0\?(\cH^1)^3}}{ (\cH^1)^2 -g^0\?g_1\? \Jpar^2 \cos^2\theta}\,,
\end{equation}
while the vectors are given by
\begin{align}
A^0 = &\? -\frac{2\? (\cH^1) ^3- g^0\? \Jpar^2\? (\cH_0\? g^0 + 3\? \cH^1 g_1)\? \cos^2 \theta}{\ex^{3\?\psi}\?\ex^{-4 U}}\,(dt+\omega ) 
\nn\\[6pt]
&\? - g^0\? r\? \left( dt + \Jpar\? (-4\? g^0 g_1^3)^{1/4} \? \sin^2\theta \, d\phi \right)
- \Jpar \?g^0 \?( h_0\? g^0 + 3\? h^1 g_1)\? \sin^2\theta \, d\phi  \,,
\end{align}
and 
\begin{align}
A^1 = \?&\? \Jpar\? \cos \theta \?\frac{\cH^1 \? (\cH_0 g^0 - \cH^1 g_1) + 2\?g^0 g_1^2 \, \Jpar^2\? \cos^2 \theta}{\ex^{3\?\psi}\?\ex^{-4 U}}\,(dt+\omega )
\nn\\[6pt]
&\? - \left( p^1 + 2\?g^0 g_1^2 \, \Jpar^2\?\sin^2\theta \right)\?\cos \theta \, d\phi \,,
\end{align}
where we have chosen the integration constants for the spatial components of the gauge fields such that they vanish asymptotically and we have used \eqref{eq:ex1-attr}.

We now can compute the conserved charges applying the procedure described in the previous section. Using the formulas for the angular momentum \eqref{eq:J-general} we find
\begin{equation}\label{J}
\mathcal{J} = \Jpar \? (g^0 h_0 + g_1 h^1)^3\,,
\end{equation}
which upon using the solution to the attractor equations for the $h$'s in terms of the charges in \eqref{eq:ex1-attr-sol}, leads to 
\begin{equation}
\mathcal{J} = \Jpar \? \left(  1 + 4\? g_1 p^1 \right)^{3/2}\,,
\end{equation}
which is indeed linear in the parameter $\Jpar$.
Using this result in the entropy formula \eqref{eq:ex1-S}, one finds the following expression for the entropy in terms of conserved quantities only
\begin{eqnarray} \label{entropy_bigJ}
S_{BH} 
= \pi  \sqrt{\frac{\sqrt{(1 + 12\? g_1\? p^1 )(1 + 4\?g_1\? p^1)^3 - 4\?\mathcal{J}^2 l_{AdS}^{-4}} - (24\? (g_1 \? p^1)^2+12\? g_1\? p^1+1) }{2\? l_{AdS}^{-4}}}\,.
\end{eqnarray}
Finally, we record the following expression for the product of the areas of the four horizons of this solution\footnote{There are in principle four (complex) roots for the warp factor of an AdS black hole solution. Our configuration is extremal, and given the form of the warp factor the roots are pairwise equal $r_{1,2} =0$ and $r_{3,4} = -\frac{-g^0 h_0 +3\? g_1 h^1}{\sqrt{2} f_0^{1/4} g^{3/4}}$. The four areas correspond to these four roots.}:
\begin{equation}
\prod_{\alpha=1}^4 A_{\alpha} 
= (4 \pi)^4 l_{AdS}^4 \left(4  q_0 (p^1)^3 + \mathcal{J}^2 \right)\,,
\end{equation}
which turns out to depend only on the cosmological constant, quantized charges and angular momentum, as for other known classes of AdS black holes \cite{Cvetic:2010mn,Castro:2012av,Toldo:2012ec,Klemm:2012vm}.

\section{The BPS entropy function}
\label{sec:entropy}

In this section, we turn to the definition of a BPS entropy function associated to our black hole solutions. Such objects have a long history, originating in the context of static BPS black holes in the ungauged theory, wherein it was observed that the attractor equations can be obtained by extremization of the central charge with respect to the scalars, while the entropy is given by its value at the extremum \cite{Ferrara:1996dd, Ferrara:1997tw}. Subsequently, it was shown that an extremization principle exists for any extremal black hole attractor, static or stationary, due to the scaling symmetry of the near horizon region \cite{Sen:2005wa, Astefanesei:2006dd}, reducing to the BPS entropy function when supersymmetric constraints are imposed. 
More recently, the corresponding BPS entropy functions for static black holes in gauged supergravity \cite{Cacciatori:2009iz, Dall'Agata:2010gj, Halmagyi:2014qza} have been matched with the topologically twisted index of the dual field theory, once it is extremized with respect to the fugacities \cite{Benini:2015eyy}.

In this paper, we have constructed rotating BPS black holes in abelian gauged supergravity, generalizing the static ones described in \cite{Cacciatori:2009iz, Dall'Agata:2010gj}, for which a corresponding BPS entropy function is expected to exist. To this end, one may attempt to apply the two techniques used to derive a BPS entropy function in the static case, namely to either ``integrate'' the BPS equations to obtain a corresponding action, or to impose BPS constraints on the general entropy function obtained by reducing the Lagrangian at the attractor region \cite{Astefanesei:2006dd}. In practice, a naive attempt to obtain an action principle for the BPS attractor equations \eqref{eq:d-scalars}-\eqref{eq:omega-eqn} would lead to to an entropy functional  rather than an entropy function, since the dependence of the various fields on the horizon coordinate must be incorporated. However, the analysis of \cite{Astefanesei:2006dd} shows that for the case of a compact horizon, such a functional should always be possible to reduce to a boundary term evaluated at the two poles.

Based on these considerations, we propose a BPS entropy function for the black hole solutions of section \ref{sec:flow}, given by the following expression 
\begin{equation}\label{eq:entropy-fun-Sen}
 \cS = -\frac\pi4\?\frac{\Iprod{\Gamma}{ I_4^\prime(\cH_0 + \Jpar\ G \cos\theta) } - 4 \? \gamma\ \Jpar^2\ \cos^2\theta}{\Iprod{G}{\cH_0}\?\sqrt{I_4(\cH_0 + \Jpar\ G \cos\theta)}}\? \cos\theta \, \Biggr|_{\,\theta = 0}^{\,\theta = \pi}\,,
\end{equation}
which is equivalent to
\begin{equation}\label{eq:entropy-fun-extremization}
 \cS = \frac{\pi}{4 \Iprod{G}{\cH_0} \sqrt{\cW(\Jpar)} } \left(\frac13  I_4(\Gamma, \cH_0, \cH_0, \cH_0)+\Jpar^2 (I_4(\Gamma, \cH_0, G, G) - 8 \gamma) \right)\,,
\end{equation}
where we used the definition \eqref{eq:BHentropy-horizon} to shorten the notation. Here we used that the gauging $G$ and the vector of parameters $\cH_0$ are such that the NUT charge \eqref{eq:NUT-ch} vanishes, while $\Jpar$ is viewed as function of the $\cH_0$ as $\Jpar(\cJ,\cH_0)$, whose explicit form is given by the expression for the angular momentum \eqref{eq:J-general}, as
\begin{equation}\label{eq:gamma-def}
 \Jpar(\cJ,\cH_0) = - 2\?\left( \Iprod{I_4^\prime(G)}{I_4^\prime(\cH_0)} - \frac12 I_4(\cH_0, \cH_0, G, G)\?\Iprod{G}{\cH_0} \rule[.1cm]{0pt}{\baselineskip}\right)^{-1} \cJ\, \equiv - \frac{\cJ}{\gamma}.
\end{equation}
Note that the expressions \eqref{eq:entropy-fun-Sen} and \eqref{eq:entropy-fun-extremization} are formally equivalent. We have shown them separately as they serve two slightly different purposes. We propose \eqref{eq:entropy-fun-Sen} as the expression for Sen's entropy function in the rotating cases we consider, where one expects a covariant angular dependent expression evaluated on the poles. On the other hand, we propose \eqref{eq:entropy-fun-extremization} as the extremization principle for the black holes that naturally generalizes\footnote{In the $I_4$-formulation we use here, the relevant extremization principle in the static case can be derived directly from \eqref{eq:H-0} by contracting the left-hand side with $I_4'(\cH_0)$ and dividing by $\Iprod{G}{\cH_0} \sqrt{\cW}$, which produces the entropy on the right hand side. If we repeat the exact same steps with the attractor equation \eqref{eq:attr-fin}, we produce the entropy together with the two other terms in \eqref{eq:entropy-fun-extremization}. This derivation bypasses \eqref{eq:entropy-fun-Sen}.} the static result of \cite{Cacciatori:2009iz, Dall'Agata:2010gj}.

It is straightforward, if cumbersome, to verify that the variation of \eqref{eq:entropy-fun-extremization} with respect to $\cH_0$ and imposing the constraint \eqref{eq:flow-conds} leads to a set of equations that are solved upon using the attractor equation \eqref{eq:attr-fin-2}, while the constraint on $\cH_0$ imply \eqref{eq:charge-constraints}. Evaluating \eqref{eq:entropy-fun-extremization} for this extremum leads to the entropy \eqref{eq:BHentropy-horizon}, as expected. While one must still solve \eqref{eq:attr-fin-2} to obtain the $\cH_0$ explicitly in terms of the charges, such a solution exists and can be written down explicitly as explained around \eqref{eq:J-general}-\eqref{eq:attr-fin-J}.

The extremization based on \eqref{eq:entropy-fun-extremization}, where the angular momentum $\cJ$ appears explicitly, can be transformed to a more traditional entropy function that depends on a corresponding fugacity $w$ by an inverse Legendre transform. Computing the derivative of \eqref{eq:entropy-fun-extremization} with respect to $\cJ$, one finds
\begin{equation}\label{eq:w-def}
w \equiv \frac{1}{\pi} \frac{\partial \cS}{\partial \cJ} = \frac{\Jpar}{\Iprod{G}{\cH_0}\?\sqrt{I_4(\cH_0 + \Jpar\? G )}} \,,
\end{equation}
i.e. it reproduces the off-diagonal $\phi$-$\tau$ term in the metric \eqref{eq:metr-Sen}. This is the object dual to the angular momentum $\cJ$ in the entropy function analysis of \cite{Astefanesei:2006dd}. One may therefore define a new BPS entropy function
\begin{equation}
 {\sf L}(\cH_0, w) \equiv \cS -\pi w\? \cJ \, \, \rule[-.2cm]{.5pt}{.8cm}_{\,\cJ = \cJ(w) } \,,
\end{equation}
where $\cJ$ is understood as a function of $w$ by inverting \eqref{eq:w-def}.
Explicitly, we find 
\begin{equation}\label{eq:free-energy-J}
{\sf L} =  \frac{\pi}{4 \Iprod{G}{\cH_0} \sqrt{\cW(\tilde\Jpar)} } \left(\frac13  I_4(\Gamma, \cH_0, \cH_0, \cH_0)+\tilde\Jpar^2 ( I_4(\Gamma, \cH_0, G, G) - 4 \gamma) \right)\,,
\end{equation}
where 
\begin{align} 
\tilde\Jpar(w,\cH_0) =&\? \Jpar(\cJ(w),\cH_0) 
\nonumber\\
=&\? \frac{1}{2\?\sqrt{I_4(G)}\?w}\?\left( \sqrt{\frac{\pi^2}{{\Iprod{G}{\cH_0}^2}} -\frac{w^2}{2}\?I_4(\cH_0,\cH_0,G,G) + 2\? \sqrt{I_4(\cH_0)\?I_4(G)}\?w^2} \right.
\nonumber\\
&\?\hspace{1.9cm} \left. - \sqrt{\frac{\pi^2}{{\Iprod{G}{\cH_0}^2}} -\frac{w^2}{2}\?I_4(\cH_0,\cH_0,G,G) - 2\? \sqrt{I_4(\cH_0)\?I_4(G)}\?w^2} \right)\,,
\end{align}
and we note that \eqref{eq:w-def} implies that
\begin{equation}
 \cW (\tilde{\Jpar}) = \frac{\tilde\Jpar^2}{\Iprod{G}{\cH_0}^2\?w^2} \,.
\end{equation}
The entropy function
\begin{equation}
 \cE = {\sf L} + \pi w\? \cJ\,,
\end{equation}
is then of the standard type, where one must extremize with respect to the scalars $\cH_0$ and $w$ to obtain the attractor equations and the expression for the angular momentum.

\subsection{The asymptotically AdS\texorpdfstring{$_4$}{4} example}
We now turn to an explicit example, namely that of the STU model in the frame \eqref{eq:STU-root}-\eqref{I4-ch-root}, with a purely electric gauging given by
\begin{equation}\label{eq:root-G}
G = \{ 0,\?0,\?0,\?0,\? g_0,\? g_1,\? g_2,\? g_3 \}^{\rm T}\,,
\end{equation}
leading to the quantities
\begin{equation}\label{eq:root-I4G}
I_4(G)= 4\? g_0 g_1g_2g_3 \,, \qquad I^\prime_4(G) = 4\?\{\? g_1 g_2 g_3,\? g_0 g_2 g_3 ,\? g_0 g_1 g_3,\? g_0 g_1 g_2, 0,0,0,0\}^{\rm T}\,.
\end{equation}
and the cosmological constant $\Lambda = -3 \sqrt{I_4(G)}= -3 \sqrt{4\? g_0 g_1g_2g_3}$, which is real and negative when all $g_I>0$ in \eqref{eq:root-G}. In addition, we are interested in a purely magnetic charge vector $\Gamma$, and therefore we can choose the vector of constants $\cH_0$ as purely magnetic, so that 
\begin{align}\label{eq:root-Gamma-H}
\Gamma = &\? \{\? p^0,\? p^1,\? p^2,\? p^3 ,\? 0,\?0,\?0,\?0 \?\}^{\rm T}\,,
\\
\cH_0 = &\? \{\? h^0,\? h^1,\? h^2,\? h^3 ,\? 0,\?0,\?0,\?0 \?\}^{\rm T}\,,
\end{align}
and the constraint \eqref{eq:final-sph-2} reads
\begin{equation}
  \Iprod{G}{\Gamma} = \sum_I g_I \? p^I = -1 \,.
\end{equation}
Note that the solution based on the above choices is related to the example discussed in section \ref{Model_ads_vac} and section \ref{sec:example-flow}, by the inverse of \eqref{eq:STU-sym-rot}. However, in this section we prefer to move to this frame, as it allows for a lift to M-theory on S$^7$ \cite{Cvetic:1999xp} and a direct connection to the dual field theory.
The objects appearing in \eqref{eq:entropy-fun-extremization} in this frame are then given by
\begin{align}
\Iprod{G}{\cH_0} = &\? \sum_I g_I \? h^I\,, \qquad 
\\
I_4(\cH_0 \pm \frac\cJ{\gamma} G)=&\, 4\,h^0 h^1 h^2 h^3 - (h^0 g_0 - h^i g_i)^2\?\frac{\cJ^2}{\gamma^2} + 4\,g_0 g_1 g_2 g_3 \?\frac{\cJ^4}{\gamma^4}
\nonumber\\
                              &\?   + 4\? (h^1 h^2 g_1 g_2 + h^1 h^3 g_1 g_3 + h^2 h^3 g_2 g_3)\?\frac{\cJ^2}{\gamma^2}\,, 
\\
\Iprod{\Gamma}{ I_4^\prime(\cH_0 \pm \frac\cJ{\gamma} G ) } = &\, 
4\,h^0 h^1 h^2 h^3 \? \sum_I \frac{p^I}{h^I} + 2\?  \left(\Iprod{G}{\Gamma}\? \Iprod{G}{\cH_0}
-2\?\sum_I g_I^2 p^I h^I \right)\?\frac{\cJ^2}{\gamma^2}\,. 
\end{align}
and $\gamma$ is expressed as a product of three constants $\gamma_i$, for $i=1,2,3$, which will be useful below,
\begin{align}
\gamma_i = &\? \left( g_0 h^0 - g_1 h^1 - g_2 h^2 - g_3 h^3 + 2\?g_i h^i \right)\,, \\
\gamma = &\? \gamma_1\? \gamma_2\? \gamma_3\,,
\end{align}
where $\gamma$ is in agreement with the definition \eqref{eq:gamma-def}. This entropy function looks somewhat complicated, but simplifies in an expansion around the static limit, $\cJ\,, w \rightarrow 0$, for which it collapses to
\begin{align}\label{eq:entropy-fun-J0}
 {\sf L} =&\?  \pi\?\frac{\sqrt{h^0 h^1 h^2 h^3}}{\Iprod{G}{\cH_0}}\? \sum_I \frac{p^I}{h^I}
\\
&\? - \frac{\pi\ \gamma}{4}\?\sqrt{h^0 h^1 h^2 h^3}\? \Iprod{G}{\cH_0} \? \left( 4 + \sum_{i=1}^3 \frac{1}{\gamma_i}\?  \left( \frac{p^0}{h^0} - \frac{p^1}{h^1} - \frac{p^2}{h^2} - \frac{p^3}{h^3} + 2\?\frac{p^i}{h^i} \right) \right)\?\frac{w^2}{\pi^2} + \cO(w^4) \,.  \nonumber
\end{align}
Here, the term for $w=0$ is of the form given in \cite{Benini:2015eyy}, upon imposing the constraint $\Iprod{G}{\cH_0}=1$. The latter can be imposed without loss of generality\footnote{The quantity $\Iprod{G}{\cH_0}$ must be nonzero by regularity, and is conventionally taken to be positive.} in the static case, in view of the homogeneity of \eqref{eq:entropy-fun-J0}, simply amounting to an overall rescaling of the $h^I$. However, such a is symmetry is not clear by inspection of  \eqref{eq:entropy-fun-extremization} in the general case.

\subsection{The \texorpdfstring{$I_4(G) = 0$}{I4(G) = 0} case}
There is a particular simplification in the case where the gauging vector is not generic, i.e. when it does not allow for an AdS$_4$ asymptotics. In this case we find that 
\begin{equation}
	\cW = I_4(\cH_0) - \Jpar^2\ ,
\end{equation}
and upon gauge fixing the linear combination
\begin{equation}
	\Iprod{G}{\cH_0} = 1\ ,
\end{equation}
we can easily invert the relation for $w(\Jpar)$ to find
\begin{equation}
	\tilde\Jpar = \frac{w\ \sqrt{I_4(\cH_0)}}{\sqrt{1+w^2}}\ , \qquad \Rightarrow \quad \cW = \frac{I_4(\cH_0)}{(1+w^2)}\ . 
\end{equation}
The entropy function in this case takes a particularly simple form,
\begin{equation}
	{\sf L}  = \pi\ \sqrt{1+w^2}\ \frac{I_4(\Gamma, \cH_0, \cH_0, \cH_0)}{12 \sqrt{I_4(\cH_0)}} + \pi \sqrt{I_4(\cH_0)} \frac{w^2}{4\ \sqrt{1+w^2}} (I_4(\Gamma, \cH_0, G, G)-4 \gamma)\ .
\end{equation}

In the extra special case of asymptotically flat solutions, $I_4'(G) = 0$ and we find that $I_4(\Gamma, \cH_0, G, G) = 4 \gamma = 4$, leading us to 
\begin{equation}\label{eq:Mink-EF}
	{\sf L}_{\text{Mink}}  = \pi\ \sqrt{1+w^2}\ \frac{I_4(\Gamma, \cH_0, \cH_0, \cH_0)}{12 \sqrt{I_4(\cH_0)}}\ .
\end{equation}
Note that in the static limit, $w=0$, the function \eqref{eq:Mink-EF} is simply equal to the central charge, thus reproducing the extremization for the static black holes in \cite{Cacciatori:2009iz,Dall'Agata:2010gj}. It is now easy to verify explicitly that upon extremization of ${\sf L}_{\text{Mink}} + \pi w \cJ$ with respect to $w$ and $\cH_0$ one finds back the explicit asymptotically flat solution of subsection \ref{subsec:nhg-examples}. To show we can change of variables by
\begin{equation}
	\phi^i \equiv \frac{1}{h_i}\ \sqrt{1+w^2}\ \sqrt{h^0 h_1 h_2 h_3}\ .
\end{equation}
If we further use that the gauge fixing choice $\Iprod{G}{\cH_0} = 1$ leads to $h^0 = -p^0$, we finally find
\begin{equation}
	\cE_{\text{Mink}} = - 2 \pi \frac{\phi^1 \phi^2 \phi^3}{1+w^2} + \pi q_i \phi^i  + \pi w \cJ\ ,
\end{equation}
in agreement with the answer in \cite{Gomes:2013cca} \footnote{Due to the five-dimensional language in the reference, the exact match works via Wick rotation of $w$ and $\cJ$.}.


\section*{Acknowledgements}

We would like to thank I. Bena, F. Benini, and B. Willett for stimulating discussions, N. Bobev and G. Bossard for useful comments on the draft, and S.M. Hosseini and A. Zaffaroni for discussions and for pointing out typos in previous versions of the paper. KH is supported in part by the Bulgarian NSF grant DN08/3 and the bilateral grant STC/Bulgaria-France 01/6. The work of SK is supported by the KU Leuven C1 grant ZKD1118 C16/16/005, by the Belgian Federal Science Policy Office through the Inter-University Attraction Pole P7/37, and by the COST Action MP1210 The String Theory Universe. CT acknowledges support from the NSF Grant PHY-1125915 and the Agence Nationale de la Recherche (ANR) under the grant Black-dS-String (ANR-16-CE31-0004). CT would like to thank the Simons Center for Geometry and Physics, Stony Brook University for hospitality during some steps of this paper.

\appendix

\section{Notation and conventions}
\label{sec:conventions}
In this paper we follow the notation and conventions of \cite{Bossard:2012xsa, Katmadas:2014faa}, and in this appendix we summarize the basic definitions that are useful in the main text, referring to those papers for more details.

\subsection{Field strengths and the symplectic section}
\label{app:Omega}
The gauge field strengths are naturally arranged in a symplectic vector with both electric and magnetic components, whose integral over a sphere defines the associated electromagnetic charges as
\begin{equation}\label{eq:dual-gauge}
 \mathsf{F}_{\mu\nu} = \begin{pmatrix} F_{\mu\nu}^I\\ G_{I\, \mu\nu}\end{pmatrix} 
\quad \Rightarrow \quad 
\Gamma=\begin{pmatrix} p^I\\ q_{I}\end{pmatrix}
 =\frac{1}{2\pi} \,\int_{S^2} \mathsf{F}\,.
\end{equation}
Here, $F_{\mu\nu}^I$ and $G_{I\, \mu\nu}$ are the field strengths and the dual field strengths of the gauge fields respectively, where the latter are defined by taking a derivative of the Lagrangian \eqref{Ssugra4D} or, equivalently, through \eqref{G-def}.
 
The physical scalar fields, $t^i$, which parametrize an $\nv$-dimensional special K\"ahler space, appear through the so called symplectic section $\cV$ in \eqref{eq:sym-sec-0}. 
The latter is subject to the constraints
\begin{equation} \label{eq:D-gauge}
  \Iprod{\bar{\cV}}{\cV} =  i \qquad
  \Iprod{\bar{D}_{\bar i}\?\bar{\cV}}{D_j\cV} =  -i\,g_{\?\bar{i}j} \,,
\end{equation}
with all other inner products vanishing, and is determined by the physical scalar fields $t^i=\frac{X^i}{X^0}$ up to a local $U(1)$ transformation. Here, $g_{\bar{\imath} j}$ is the K\"ahler metric and the covariant derivative $D_i\cV$ contains the K\"{a}hler connection $Q_\mu$, defined through the K\"{a}hler potential as
\begin{equation} 
Q = \Im[ \partial_ i \cK\, dt^i] \label{Kah-def}\ ,\qquad
\cK = - \mbox{ln}\left( \tfrac{i}6\, c_{ijk} (t-\bar t)^i(t-\bar t)^j(t-\bar t)^k\right) \,. 
\end{equation}

We introduce the following notation for any symplectic vector $\Gamma$
\begin{align}\label{ch-def}
Z(\Gamma) = \Iprod{\Gamma}{\cV} \,,\qquad
Z_i(\Gamma) = \Iprod{\Gamma}{D_i \cV} \,.
\end{align}
When an argument does not appear explicitly in $Z$ or $Z_i$, the vector of charges in \eqref{eq:dual-gauge} is understood, as is standard in literature. In addition, the operation is applied component wise when the argument has additional indices, e.g. when it is form valued. With these definitions one can introduce a scalar dependent complex basis for symplectic vectors, given by $(\cV,\, D_i\cV)$, so that any vector $\Gamma$ can be expanded as
\begin{equation}\label{Z-expand}
\Gamma = 2 \Im[- \bar{Z}(\Gamma)\,\cV + g^{\bar \imath j} \bar{Z}_{\bar \imath}(\Gamma)\, D_j \cV]\,,
\end{equation}
so that the symplectic inner product is computed from \eqref{eq:D-gauge} as
\begin{equation}\label{inter-prod-Z}
\Iprod{\Gamma_1}{\Gamma_2} = 2 \Im[- Z(\Gamma_1)\,\bar{Z}(\Gamma_2)
   +g^{i\bar \jmath} Z_i(\Gamma_1) \, \bar{Z}_{\bar \jmath}(\Gamma_2)]\,.
\end{equation}
Furthermore, we introduce the scalar dependent complex structure $\mathrm{J}$, defined as
\begin{equation}\label{CY-hodge}
\mathrm{J}\cV=-i \cV\,,\quad
\mathrm{J} D_i\cV=i  D_i\cV\,,
\end{equation}
which can be solved to determine $\mathrm{J}$ in terms of the period matrix $\cN_{IJ}$ in \eqref{G-def}, see e.g.~\cite{Ceresole:1995ca} for more details. With this definition, the complex self-duality condition on the gauge field strengths is given by
\begin{equation}\label{cmplx-sdual}
 \mathrm{J}\, \mathsf{F} =-* \mathsf{F} \,,
\end{equation}
which is the duality covariant form of the relation between electric and magnetic components. In addition, we mention the important relation
\begin{equation}\label{VBH-def}
 \tfrac12\,\Iprod{\Gamma}{\mathrm{J}\,\Gamma}=
 |Z(\Gamma)|^2 + g^{i\bar \jmath} Z_i(\Gamma) \, \bar{Z}_{\bar \jmath}(\Gamma)
 \equiv V_{\text{\tiny BH}}(\Gamma)\,,
\end{equation}
where we defined the black hole potential $V_{\text{\tiny BH}}(\Gamma)$.

\subsection{Identities involving the quartic invariant}
\label{app:I4}
We now summarize a number of useful relations involving the quartic invariant, defined for all symmetric models. The starting point is the definition of this invariant, $I_4(\Gamma)$, for any symplectic vector, $\Gamma$, as \cite{Ferrara:1997uz,Ferrara:2006yb}
\begin{eqnarray}
I_4(\Gamma)&=& \frac{1}{4!} t^{\sM\sN\sP\sQ}\Gamma_\sM\Gamma_\sN\Gamma_\sP\Gamma_\sQ
\nonumber\\
         &=& - (p^0 q_0 + p^i q_i)^2 + \frac{2}{3} \,q_0\,c_{ijk} p^i p^j p^k- \frac{2}{3} \,p^0\,c^{ijk} q_i q_j q_k 
             + c_{ijk}p^jp^k\,c^{ilm}q_lq_m\,, \label{I4-ch}
\end{eqnarray}
where $M$, $N$\dots are indices encompassing both electric and magnetic components and we also defined the completely symmetric tensor $t^{\sM\sN\sP\sQ}$ for later use. It is also convenient to define a symplectic vector out the first derivative of the quartic invariant, $I_4^\prime(\Gamma)$, as
\begin{equation} \label{I4-der-basis}
I_4^{\prime}(\Gamma)_{\sM} \equiv \Omega_{\sM\sN}\frac{\partial I_4(\Gamma)}{\partial \Gamma_\sN} 
 = \frac{1}{3!}\, \Omega_{\sM\sN} t^{\sN\sP\sQ\sR} \Gamma_\sP \Gamma_\sQ \Gamma_\sR \,,
\end{equation}
where $\Omega_{\sM\sN}$ is the inverse of the symplectic form $\Omega^{\sM\sN}$, so that the following relations hold
\begin{equation}
 \Iprod{\Gamma}{I^\prime_4(\Gamma)} = 4\,I_4(\Gamma)  \ , \qquad I^\prime_4(\Gamma,\Gamma,\Gamma) = 6 I_4^\prime(\Gamma) \ .
\end{equation}
Throughout this paper, all instances of $I_4(\Gamma_1,\Gamma_2,\Gamma_3,\Gamma_4)$ will denote the contraction of the tensor $t^{\sM\sN\sP\sQ}$ in \eqref{I4-ch} with the four displayed charges, without any symmetry factors, except for the case with a single argument, as in $I_4(\Gamma)$ and $I^\prime_4(\Gamma)$. For more details on this tensor, see \cite{Bossard:2013oga} in the real basis and \cite{Ferrara:2011di} in the complex basis, to be defined shortly.

We now record some identities that are used repeatedly in the main text, starting with the fundamental property
\begin{equation}
 I_4^{\prime}( I_4^{\prime}(\Gamma) ) = -16\, I_4(\Gamma)^2 \Gamma\,.
\end{equation}
Further properties include the quintic and heptic identities 
\begin{equation}\label{eq:quint}
 I_4^{\prime}( I_4^{\prime}(\Gamma), \Gamma, \Gamma ) = -8\, I_4(\Gamma) \Gamma \,,
 \qquad
 I_4^{\prime}( I_4^{\prime}(\Gamma), I_4^{\prime}(\Gamma), \Gamma ) = 8\, I_4(\Gamma)\, I_4^{\prime}(\Gamma)\,.
\end{equation}
Finally, the projection operator
\begin{equation}\label{eq:proj}
 I_4^{\prime}( I_4^{\prime}(\Gamma_1), \Gamma_1, \Gamma_2 ) 
 = 2\, \Iprod{\Gamma_1}{\Gamma_2}\,I_4^{\prime}(\Gamma_1) + 2\, \Iprod{I_4^{\prime}(\Gamma_1)}{\Gamma_2}\,\Gamma_1 \,,
\end{equation}
for any $\Gamma_1$, $\Gamma_2$, is useful in the analysis of sections \ref{subsec:nhg-BPS} and \ref{sec:flow}.

One can rewrite the quartic invariant in the complex basis \cite{Cerchiai:2009pi}, leading
to the following alternative definition
\begin{align} 
I_4(\Gamma) =&\, \left( Z \, \bar Z - Z_i \, \bar Z^i \right)^2 
- c_{mij} \bar Z^i \bar Z^j \, c^{mkl} Z_k  Z_l 
+  \frac{2}{3} \, \bar Z \, c^{ijk} Z^i Z^j Z^k  + \frac{2}{3} \, Z \,c_{ijk} \bar Z^i \bar Z^j \bar Z^k\,.
\label{I4-def}
\end{align}
Despite the appearance of the central charges, this expression is by construction independent of the scalars, which only appear due to the change of basis in \eqref{Z-expand}. We will not make use of this basis, but we do note the following identity,
\begin{eqnarray}
 \tfrac12\, I_4(\Gamma, 2 \Im\cV, 2 \Im\cV)
    &=& 8\,\Im(Z(\Gamma))\,\Im\cV
    +16\,\Re(Z(\Gamma))\,\Re\cV
    -2\,\mathrm{J}\,\Gamma
      \,,
    \label{I4toJ}
\end{eqnarray}
central to the analysis of section \ref{subsec:BPS}. We refer to \cite{Katmadas:2014faa} for its derivation. A more extensive list of useful $I_4$ identities can be also found in the appendix of \cite{Halmagyi:2014qza}.

\section{The Kunduri-Lucietti-Reall solution}
\label{sec:klr}
In this section, we present the doubly spinning black hole solution of Kunduri, Lucietti and Reall \cite{Kunduri:2006ek} in new variables, which bring the form of the solution closer to that of asymptotically flat BPS solutions and are adapted for Kaluza--Klein reduction. 

We start with the BPS equations in five-dimensional gauged supergravity coupled to $n_v$ vector multiplets, labelled by indices $i, j, k\ldots = 1,\ldots, n_v+1$. The scalars $X^i$ satisfy
the constraint:
\begin{equation}
  \label{conda}
  X_i X^i =1\,,\qquad X_i \equiv \frac 16\? c_{ijk} X^j X^k \ ,
\end{equation}
so that only $n_v$ of them are independent, while the field strengths $F^i_{\mu\nu}$ are unrestricted and include the graviphoton. The tensor $c_{ijk}$ is assumed to satisfy \eqref{eq:symm-cub}, so that we restrict to symmetric models. A general BPS solution in the so called timelike class is constructed as follows. 
Given a K\"ahler 4-manifold with a metric, $ds^2_{\text b }$, we choose the associated
K\"ahler form, $\Kah$. Given these data, the metric and gauge fields of a supersymmetric solution can be
written locally as:
\begin{eqnarray}\label{eq:bps_metric}
  ds^2&=& -f^2(dt+\omega)^2 + f^{-1}ds^2_{\text b } \nn\\
  {}F^i &=&  d(X^i e^0) + \Theta^{i} - 9\?g\?f^{-1}c^{ijk} V_j X_k\?\Kah\,.
\end{eqnarray}
Here, $e^0=f(dt+\omega)$, the $V_i$ are the FI gauging parameters, the $ \Theta^{i}$ are arbitrary closed self-dual forms on the base, and $f>0$, is assumed to be a globally defined function. A positive orientation is chosen using
$e^0\wedge \eta$ as the volume form, where $\eta$ is a positive orientation on the base.
The fluxes $ \Theta^{i}$ are constrained in terms of the Ricci form on the base, $\Ric$, as
\begin{equation}\label{eq:Riccif-con}
 3\?g\? V_i \?\Theta^{i} - 27\?g^2 f^{-1}c^{ijk} V_i V_j X_k\?\Kah = \Ric \,.
\end{equation}
Contraction with the \Kahf form leads to the scalar condition
\begin{equation}\label{eq:Riccis-con}
f\?R_s + 108\?g^2 c^{ijk} V_i V_j X_k=0\,,
\end{equation}
where $R_s$ is the Ricci scalar on the base. The last equation implies that the overall scale of the function, $f$, can be absorbed by a conformal rescaling of the base metric, $ds^2_{\text b }$, consistent with \eqref{eq:bps_metric} upon appropriate rescaling of the time coordinate and $\omega$.
The one-form, $\omega$, is determined by solving:
\begin{equation}\label{bps_omeg}
  f d\omega =G^+ + G^-, \qquad X_i  \Theta^{i}=- \frac23 G^+ \,,
\end{equation}
while the scalars are obtained by solving the Maxwell equations, as given in \cite{Gutowski:2004yv}, for the field strengths in \eqref{eq:bps_metric}.

\subsection{The KLR solution in 5d}
We now turn to the solution of \cite{Kunduri:2006ek}, whose base space is given by
\begin{equation} \label{eq:base-KLR}
 ds^2_{\text b } = \frac{dr^2}{r\?(\Delta_0 +  4\?g^2 r)} + \frac{r}{\Delta_\theta}\? d\theta^2 + r \?\Delta_\theta\?\sin^2\!\theta\? d\phi^2  
+r\?(\Delta_0 +  4\?g^2 r)\,\left( d\psi + \cos\theta\? d\phi \right)^2 \,, 
\end{equation} 
where 
\begin{align}\label{eq:base-consts}
\Delta_0 =&\, \left(1 +g^2\? \gamma +2\?\left(1 - \cosh\delta \right)  \right) \,,
\nonumber\\
\Delta_\theta =&\, \cosh\delta + \sinh\delta\? \cos\theta \,,
\end{align}
and $\gamma$, $\delta$ are constants parametrizing a deformation away from the Bergmann manifold, to which \eqref{eq:base-KLR} reduces for $\gamma=\delta=0$. The particular parameterization of the coefficient $\Delta_0$ as in \eqref{eq:base-consts} is chosen for convenience in comparing with the base space of the KLR solution. The latter can be put in the form \eqref{eq:base-KLR} by the coordinate transformation
\begin{gather}\label{eq:coo-KLR}
 \left\{\? \psi\,, \, \phi\,, \, r\,, \, \theta \,\right\} \big|_{\scriptscriptstyle KLR} = 
 \left\{\? \frac12\?\frac{\sqrt{\Xi_b}}{\sqrt{\Xi_a}}\?(\psi + \phi)\,, \,\, \frac12\?\frac{\sqrt{\Xi_a}}{\sqrt{\Xi_b}}\?(\psi - \phi)\,, \,\, 2\?\left( \sqrt{\Xi_a \? \Xi_b}\,r \right)^{1/2} , \,\, \frac\theta 2 \, \right\}\,, 
\end{gather}
where $|_{\scriptscriptstyle KLR}$ denotes the variables of the same name in \cite{Kunduri:2006ek} and $\Xi_a=1-g^2 a^2$, $\Xi_b=1-g^2 b^2$ depend only on the two rotation parameters, $a$, $b$ used in that reference. The latter are related to the parameters $\gamma$, $\delta$ as
\begin{align}\label{eq:pars-KLR}
 \cosh\delta = &\, \frac12\,\left( \frac{\sqrt{\Xi_a}}{\sqrt{\Xi_b}} + \frac{\sqrt{\Xi_b}}{\sqrt{\Xi_a}} \right)\,, 
 \nonumber \\ 
 \gamma = &\, \frac{1}{\sqrt{\Xi_a \? \Xi_b}}\left( 2\? \left( \tfrac{a+b}g + a\?b \right) +\frac3{g^2}\?\left( 1- \sqrt{\Xi_a \? \Xi_b} \right) \right)\,.
\end{align}
In the rest of this appendix we present the doubly spinning black hole solution constructed over the base \eqref{eq:base-KLR}, noting that \eqref{eq:coo-KLR}-\eqref{eq:pars-KLR} can be used to transform the various fields of this solution to the ones in \cite{Kunduri:2006ek}.

The K\"ahler form associated to the metric \eqref{eq:base-KLR} is
\begin{equation}\label{eq:Kah-KLR}
\Kah = dr\wedge\left( d\psi + \cos\theta\? d\phi \right) -r\?\sin\theta\? d\theta\wedge d\phi\,,
\end{equation}
which is manifestly anti-selfdual, while the resulting Ricci form and Ricci scalar read
\begin{align}\label{eq:Ricci-base-g}
\Ric =&\, - 6\?g^2\? \Kah + \left( \? g^2\gamma +3\?(1-\Delta_\theta) \right)  \?\sin\theta\? d\theta\wedge d\phi\,,
\\
R_s =&\, -24\?g^2 -\frac{2}{r}\? \left( \? g^2\gamma +3\?(1-\Delta_\theta) \right) \,.
\end{align}
Note that the Ricci form differs from the K\"ahler form only by components along the $S^2$ part of the metric \eqref{eq:base-KLR} and that \eqref{eq:Ricci-base-g} reduces to a K\"ahler--Einstein condition asymptotically. With this decomposition, the conditions \eqref{eq:Riccif-con}-\eqref{eq:Riccis-con} reduce to
\begin{gather}
 3\?g\? V_i \?\Theta^{i} = \frac{1}{2\?r}\? \left( \? g^2\gamma +3\?(1-\Delta_\theta) \right)\? \Theta_0 \,,
 \\
108\?g^2 \?f^{-1}c^{ijk} V_i V_j X_k= 24\?g^2 + \frac{2}{r}\? \left( \? g^2\gamma +3\?(1-\Delta_\theta) \right) \,,
\end{gather}
where 
\begin{equation}
 \Theta_0 = dr\wedge\left( d\psi + \cos\theta\? d\phi \right) + r\?\sin\theta\? d\theta\wedge d\phi \,.
\end{equation}

Given that the gauging vector $V_i$ is assumed to have maximal rank and a single two-form is expected to arise for a single centre solution (in this case $\Theta_0$), these equations can be solved by an ansatz linear in $1/r$ for both the $\Theta^{i}$ and $f^{-1} X_i$, leading to
\begin{align}
 &\?f^{-1} X_i = \frac{V_i}{[V]^3}\?\left( 1 - \frac{\Delta_\theta-1}{4\?g^2\?r} \right) + \frac{\mu_i}{12\?r}\,, \label{eq:f-sol}
 \\
 \Theta^i =&\? \frac{9}{2}\?c^{ijk}V_j\?\left( g\? \frac{\mu_k}{3\?r} -  \frac{V_k}{[V]^3}\?\frac{\Delta_\theta-1}{g\? r} \right)\?\Theta_0\,, \label{eq:TH-sol}
\end{align}
Here, we use the shorthand
\begin{equation}
 [V]^3 = \frac92\?c^{ijk} V_i V_j V_k\,.
\end{equation}
and $\mu_i$ is a vector of arbitrary constants\footnote{In principle, one can consider two different vectors $\mu_i$ in \eqref{eq:f-sol} and \eqref{eq:TH-sol}, both subject to \eqref{eq:gaTmu}, but the remaining BPS equations imply that they must be equal.} subject to the constraint
\begin{equation}\label{eq:gaTmu}
 \gamma =  \frac{9}{2}\?c^{ijk}V_i V_j \mu_k\,,
\end{equation}
which we will use to eliminate $\gamma$ in what follows, in favor of unconstrained $\mu_i$. The final equation to solve is \eqref{bps_omeg}, i.e.
\begin{equation}
d\omega^+ = -\frac32\, f^{-1} X_i \Theta^i \,,
\end{equation}
which can be solved for $\omega$ using \eqref{eq:f-sol}-\eqref{eq:TH-sol}, by
\begin{align}
 \omega =&\? \left( -2\?g\?r + \frac{3\?g}{32\?r}\?c^{ijk}V_i \mu_j \mu_k \right)\?\left( d\psi + \cos\theta\? d\phi \right) 
 \nonumber\\
&\? +\frac{1}{2\?g\?[V]^3}\?\left[ \left( \Delta_0\?\left(1 - \frac{\Delta_\theta-1}{4\?g^2\?r} \right) - \Delta_\theta + \frac{\cosh{\delta}-1}{2\?g^2\?r} \right)\?\left( d\psi + \cos\theta\? d\phi \right) \right. 
 \nonumber\\
&\? \qquad \qquad \qquad \quad \left. +\left(1 - \frac{\Delta_\theta-1}{4\?g^2\?r} \right) \? \sinh{\delta}\?\sin^2{\theta}\,d\phi  \right]\,,
\end{align}
where the asymptotic $\cO(r)$ term was chosen for asymptotic regularity, following \cite{Kunduri:2006ek}. This completes the doubly spinning black hole solution, as one can show that the Maxwell equations are automatically satisfied.

\subsection{The KLR solution in 4d}

The five-dimensional solution described above can be straightforwardly reduced along $\psi$ to obtain a solution to four dimensional abelian gauged supergravity. The resulting geometry is not asymptotically AdS$_4$, since the five dimensional theory is electrically gauged and dimensional reduction of the theory described above leads to a prepotential of the type \eqref{prep-def-0}, which only has AdS$_4$ vacua for a mixed gauging. The relevant gauging vector $G$ in the four dimensional theory is rank-3, i.e. of the type \eqref{eq:rank3-vec}, so that the solution asymptotes to the hvLif geometry \eqref{eq:metr-rank-3}. In the special case of equal rotation, $\delta=0$, the reduction to four dimensions was performed in \cite{Hosseini:2017mds}, and here we extend this result to the more general case of arbitrary $\delta$.

Before presenting the result in four dimensions, we point out a subtlety related to the periodicity of the coordinate $\psi$ in \eqref{eq:base-KLR}, along which the reduction is performed. As explained in \cite{Gutowski:2004yv}, the K\"ahler base of a BPS solution must admit an SU(2) structure $\Kah^\alpha$, $\alpha=1,2,3$, such that 
\begin{equation}
 d \Kah^{(1)} = 0\,, \qquad d \Kah^{(2)} = P \wedge \Kah^{(3)}\,, \qquad d \Kah^{(3)} = - P \wedge \Kah^{(2)}\,,
\end{equation}
where $\Kah^{(1)}=\Kah$ is the \Kahf form and $P$ is the potential for the Ricci form, $R =d P$. One can explicitly check that for \eqref{eq:base-KLR} with the \Kahf form in \eqref{eq:Kah-KLR}, the remaining two forms are
\begin{align}
 \Kah^{(2)} = \frac{dr}{\sqrt{\Delta_0 +  4\?g^2 r}}\wedge\left( \sin{(\cosh\delta\?\psi)}\?\frac{d\theta}{\sqrt{\Delta_\theta}} - \cos{(\cosh\delta\?\psi)}\? \sqrt{\Delta_\theta}\?\sin\!\theta\? d\phi \right)\,,
 \nonumber\\
 \Kah^{(3)} = \frac{dr}{\sqrt{\Delta_0 +  4\?g^2 r}}\wedge\left( \sin{(\cosh\delta\?\psi)}\?\frac{d\theta}{\sqrt{\Delta_\theta}} + \cos{(\cosh\delta\?\psi)}\? \sqrt{\Delta_\theta}\?\sin\!\theta\? d\phi \right)\,,
 \end{align}
so that the Hopf fiber does not have unit charge. This can be easily remedied by performing the rescaling 
\begin{equation}
\psi \rightarrow \frac{\psi}{\cosh\delta} \,, \quad \phi \rightarrow \frac{\phi}{\cosh\delta}\,, 
\end{equation}
so that the U(1) fiber $d\psi + \cos{\theta}\?d\phi$ is uniformly rescaled. This is crucial for reproducing the four dimensional Dirac quantization condition \eqref{eq:final-charge-constr} and the asymptotic value of the metric function $\Delta$ in \eqref{eq:Delta-sol}.\footnote{Note that this rescaling can be applied already at the level of the base space \eqref{eq:base-KLR}, before building the 5d solution on top of it, but we prefer to give the explicit connection with the form the solution was given in \cite{Kunduri:2006ek}.}

The dimensional reduction is otherwise completely straightforward and we simply present some of the results here. The metric is as in \eqref{eq:metr-bps}, with base space as in \eqref{eq:3d-metr} where $\ex^\psi$ is given as
\begin{equation}
 \ex^{2\?\psi} = \frac{1}{\cosh\delta}\?(\Delta_0 +  4\?g^2 r)\? r^2 \,,
\end{equation}
with $\Delta_0$ as defined in \eqref{eq:base-consts} and \eqref{eq:gaTmu}. This is consistent with \eqref{eq:metr-rank-3}, as anticipated above. The metric along the sphere is as in \eqref{eq:base-bps-sph} with 
\begin{equation}
 \Delta(\theta) = 1 + \tanh\delta\?\cos\theta\,.
\end{equation}
The symplectic section is determined by its imaginary part, as
\begin{align}\label{eq:KLR-sol}
 \ex^{2\?\psi} \? \sv =&\? \ex^{2\psi} \? \frac{\cosh{\delta} - \sinh{\delta}\?\cos{\theta}}{32\?g^4} \? I_4^\prime(G)   
\nonumber\\
&\? + \frac{r^2}{4\?g}\,\left(\frac{1}{2}\?I_4^\prime(G,G,\Gamma) - \frac{\cosh{\delta}}{8\?g^3}\?\left( \frac{1}{4}\?I_4^\prime(G,G,\Gamma,\Gamma) + \tanh^2{\delta} \right) \? I_4^\prime(G) \right)
\nonumber\\
&\? + \sqrt{\frac{\Delta_0}{\cosh\delta}}\, \cH_0 \, r - \frac{\tanh{\delta}}{8\?g^3}\?r\?(\Delta_0 +  4\?g^2 r) \?\cos{\theta}\? G\,,
\end{align}
where $\cH_0$ is a vector satisfying the condition \eqref{eq:attr-fin}, i.e.
\begin{align} \label{eq:KLR-J}
  \Gamma = &\? \frac{1}{4}\? I^\prime_4\left(\cH_0, \cH_0, G \right) + \frac{1}{2}\? \Jpar^2\? I^\prime_4\left( G \right)\,,
\nn\\
 \Jpar = &\? -\frac{\sinh{\delta}}{8\?g^3}\? \sqrt{\frac{\Delta_0}{\cosh\delta}} \,,
\end{align}
and we defined the parameter $\Jpar$ for this solution.
Restricting to the STU model for simplicity, the gauging vector in the four dimensional theory is purely electric and given by
\begin{equation}\label{eq:KLR-G}
 G = \sqrt{2}\?\{ 0,0,0,0,\cosh{\delta},g ,g ,g \}^{\rm T}\,,
\end{equation}
which is a vector of rank-3, as mentioned above. Finally, the components of the charge vector $\Gamma = \{ p^I\,, q_I \}$ are given by
\begin{align} \label{eq:KLR-charge}
 p^I = &\?-\? \frac1{ \sqrt{2}\?\cosh{\delta} }\? \{ 1, 0, 0, 0 \}\,, 
 \nn\\
 q_0 = &\? \frac1{8\?\sqrt{2}\?g^3}\?\left( \frac2{\cosh\delta}\?\prod_i(1+g^2\?\mu_i) - \sum_i(1+g^2\?\mu_{i})(1+g^2\?\mu_{i+1}) +1 \right)  \,,
 \nn\\
 q_i = &\? \frac1{8\?\sqrt{2}\?\cosh\delta}\?\left( 2\? g^2\?\mu_{i+1}\?\mu_{i+2}- g^2\?\sum_i \mu_{i} \?\mu_{i+1} -2\?\mu_i \right)  \,,
\end{align}
where we use the cyclic convention mod $3$ for the indices $i,j,k$. Given this charge vector, the explicit expression for $\cH_0$ in the STU model is given in \eqref{eq:ex2-sol}.



\bibliography{AdSrot} \bibliographystyle{JHEP}


\end{document}